\documentclass[journal, 10]{IEEEtran}
\usepackage{amsmath,amsfonts}
\usepackage{algorithmic}
\usepackage{multirow}
\usepackage{algorithm}
\usepackage{listings,multicol}
\usepackage{array}
\usepackage{textcomp}
\usepackage{stfloats}
\usepackage{balance}
\usepackage{url}
\usepackage{amssymb}
\usepackage{pifont}
\usepackage{verbatim}
\usepackage{graphicx}
\usepackage{cite}
\usepackage[table]{xcolor}  
\newcommand{\cmark}{\ding{51}}%
\newcommand{\xmark}{\ding{55}}%
\definecolor{mycolor}{rgb}{1,0.0,0.5}
\newcommand{\remark}[1]{}
\def\nm{\textsf{CN2F}}

\usepackage{subcaption}
\usepackage{url}

\usepackage{flushend}
\usepackage{tikz}
\newcommand*\circled[1]{\tikz[baseline=(char.base)]{
            \node[shape=circle,draw,inner sep=0.25pt] (char) {#1};}}
\usepackage[hidelinks]{hyperref}
\begin{document}

\title{\nm: A Cloud-Native Cellular Network Framework}

\author{Sepehr Ganji$^1$, Shirin Behnaminia$^1$, Ali Ahangarpour$^1$, Erfan Mazaheri$^1$, Sara Baradaran$^1$,  Zeinab Zali$^1$, Mohammad Reza Heidarpour$^1$, Ali Rakhshan$^2$, Mahsa Faraji Shoyari$^2$\\
$^1$Department of Electrical and Computer Engineering, Isfahan University of Technology, Isfahan, Iran\\
$^2$Mobile Communications Company Research and Development Center, Tehran, Iran
}

\maketitle

\begin{abstract}
Upcoming cellular networks aim to improve the efficiency and flexibility of mobile networks by incorporating various technologies, such as Software-Defined Networking (SDN), Network Function Virtualization (NFV), and Network Slicing (NS). There exist open-source projects that implement components of different cellular generations. In this paper, we elaborate on how to use these open-source projects to realize a flexible and extendable testbed for conducting experiments on the future generation of cellular networks. In particular, a Cloud-Native Cellular Network Framework (\nm{}) is presented, which uses \textit{OpenAirInterface}'s codebase to generate cellular Virtual Network Functions (VNFs) and deploys \textit{Kubernetes} to disperse and manage them among multiple worker nodes. Moreover, \nm{} leverages \textit{ONOS} and \textit{Mininet} to emulate the effect of the IP transport networks in the fronthaul and backhaul of real-world cellular networks. Using \nm{}, we implement different network scenarios, including Edge Computing (EC), Cloud Computing (CC), and Radio Access Network (RAN) slicing, to showcase the effectiveness of the proposed testbed for academia and industrial Research and Development (R\&D) activities.
\end{abstract}


\section{Introduction}
The fifth generation of mobile communication networks (5G) promises the support of a range of applications from Ultra-Reliable Low-Latency Communication (URLLC) to enhanced Mobile Broadband (eMBB) to massive Machine-Type Communication (mMTC) connections. The diversity of supported applications hinders to use of the conventional one-size-fits-all structure for future cellular networks, thereby necessitating innovative models. In order to increase flexibility and efficient resource sharing among different application sectors, new models are supposed to be built upon \textit{virtualization/softwarization} technologies, such as Software-Defined Networking (SDN), Network Function Virtualization (NFV), and Network Slicing (NS), among others \cite{ref21}.

In recent years, several AI/ML-based algorithms have been proposed to enable intelligent and autonomous network management, which should be tested and evaluated before their commercial rollout \cite{ref1, ref2, ref3, ref6, ref7, 10449568, FOWDUR2023106032}. These algorithms use a large amount of data extractable from both the Radio Access Network (RAN) and Core Network (CN) to learn patterns and automatically enhance network operations \cite{10.1145/3610402, 9647461}. Small-scale testing of such algorithms allows developers to identify potential issues and address problems before a large-scale rollout. In this way, different cellular network testbeds are designed, which enable researchers to evaluate real-world network scenarios \cite{ref4, ref5}.

\begin{figure}[t]
    \centering{\includegraphics[width=0.48\textwidth]{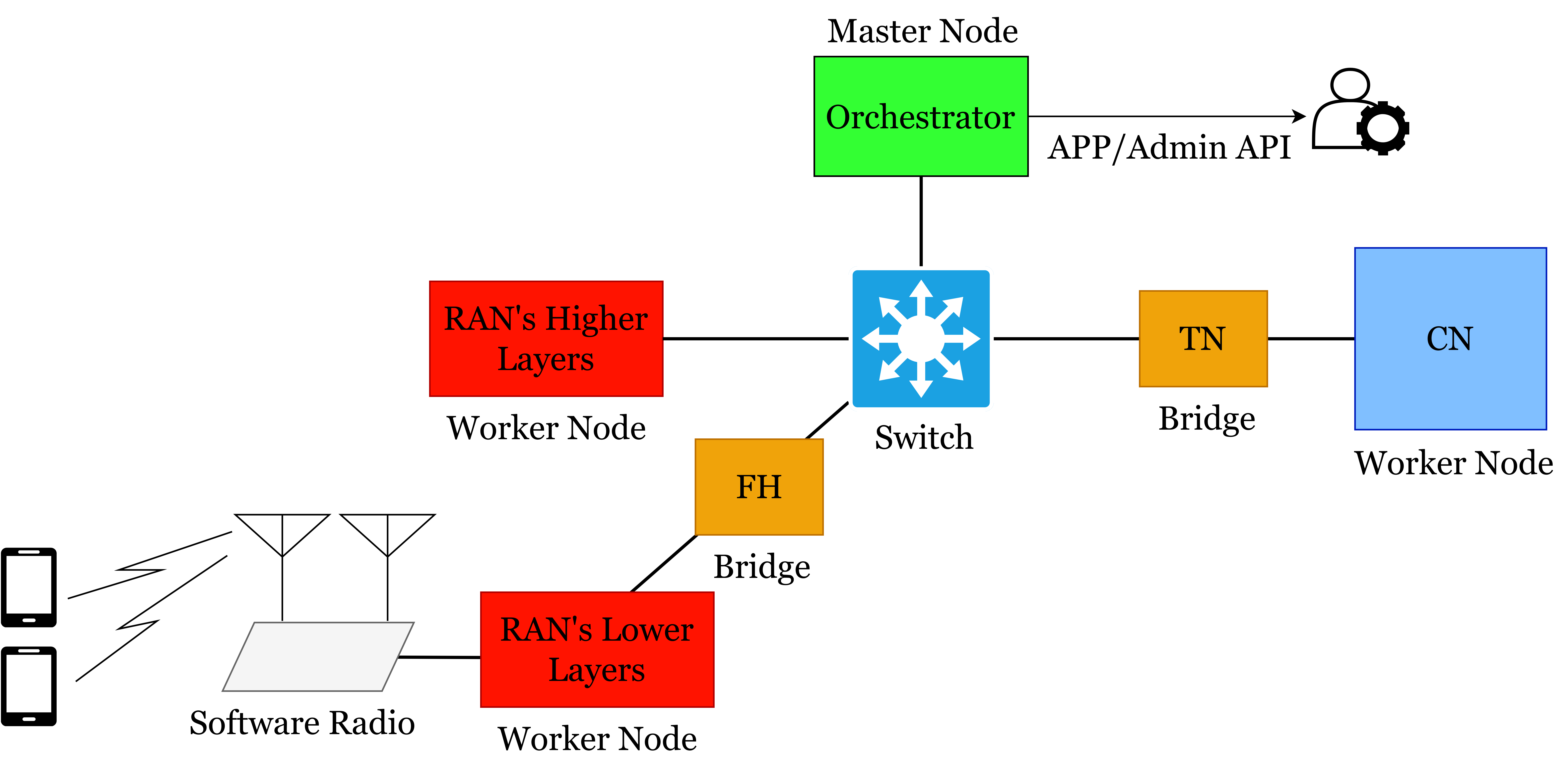}}
	\caption{\nm{} architecture with a master node and three worker nodes\label{fig:fig1}}
\end{figure}

In this paper, we propose the \nm{}, a simple Cloud-Native Cellular Network Framework, as a general framework to build prototypes for future generations of cellular networks. Fig.~\ref{fig:fig1} depicts the \nm{} structure. The \nm{} comprises a cluster of four nodes (one master and three workers), an L2/L3 switch, two bridge nodes, and a software-defined network.

First, we review the main technological concepts and frameworks such as containers, Docker \cite{docker}, Kubernetes \cite{kuber}, Mininet \cite{mininet}, and RAN splitting. Then, we explain how to use these primitives to set up the \nm{} framework (including the cluster and bridges). Specifically, for the \nm{} setup, we provide an \textit{Ansible playbook} to prepare and install necessary packages on nodes, which is a principled method to make a Kubernetes cluster. We also describe how to create cellular Virtual Network Functions (VNFs) (in the form of \textit{pods}\footnote{A Pod is a group of one or more containers, with shared storage and network resources, and a specification for how to run the containers.} in Kubernete) from \textit{Docker Images}. For the cellular VNFs, we take the OpenAirInterface (OAI) project as one of the existing open-source candidates for the software implementation of the 4G/5G RAN and CN components (e.g., MME, HSS, etc., for 4G and AMF, UDM, SMF, etc., for 5G). Finally, we demonstrate using \nm{} to investigate the importance of VNF placement and RAN slicing as two significant capabilities of future cellular networks.

The results of this paper are reproducible, and the source codes alongside scripts to set up the \nm{} and execute the use cases are publicly available on our GitHub\footnote{Available at \href{https://github.com/CN2F/core}{https://github.com/CN2F/core}} repository for the academia and industrial community.

The remainder of this paper is structured as follows. Section \ref{Sec2} provides the necessary background and definitions. Section \ref{Sec3} describes the \nm{} structure in detail. Section \ref{Sec4} explains our evaluation and the implementation of two use cases in our framework. Section \ref{Sec5} reviews alternative testbeds proposed in the literature, along with their goals and applications. Finally, Section \ref{Sec6} concludes the paper.

\section{Background}\label{Sec2}
In this section, we briefly review the required background material for building a cloud-native cellular network framework. First, we discuss topics specific to cellular networks, such as different generations of cellular networks along with available options and open-source projects usable for building and deploying cellular VNFs on a given testbed/cluster. Then, in the second part, we cover primitive technologies needed at the time of generating cloud-native applications and building a cluster over which a cloud-native application is deployed. Though the topics elaborated on in the second part are general, our objective is to show how to use these technologies to achieve a flexible cloud-native cellular network. 

\begin{figure*}[t]
	\centering{\includegraphics[width=0.8\textwidth]{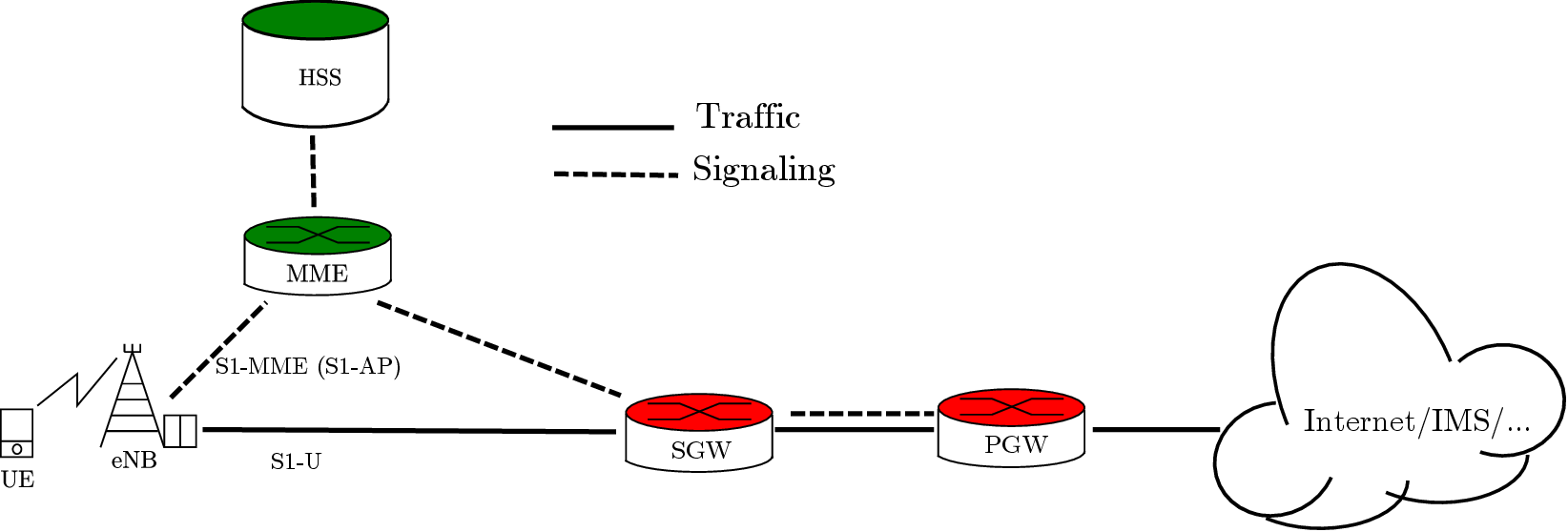}}
	\caption{A simplified LTE architecture \label{fig:lte}}
\end{figure*} 

\subsection{Cellular Network Concepts}
Wireless mobile networks first started as an extension to the Public Switched Telephone Networks (PSTN) in order to support mobile users. In the beginning, they were composed of a single high-power base station to cover an area of around 70 to 80 km radius. To increase the capacity (number of subscribers), the \textit{cellular idea} emerged. The cellular structure suggests using multiple low-power base stations through which the whole area is tiled by (cell-like) hexagons and the frequency reuse becomes possible. The downside of the cellular structure is the need for handoff and interference management. However, because of its capacity-increasing benefits, the underlying structure of all current Wireless Wide Area Networks (WWAN) obeys the cellular pattern.

\subsubsection{Cellular Generations and Open-Source Projects}
\paragraph{1G and 2G Networks}
The first generation (1G) was analog and only supported voice. The 2G is characterized by being the first that used digital modulation, and hence, featured higher service quality (e.g., encoding), security (e.g., encryption), new services (e.g., short message service), and more efficient Radio Frequency (RF) spectrum usage through Time Division Multiple Access (TDMA) and Code Division Multiple Access (CDMA). The dominant standard of the 2G network is GSM (Global System for Mobile communication), which is still alive due to the vast investment and capability to support Internet of Things (IoT) applications. For the open-source software projects that implement 2G network components, we can refer to OpenBTS and YateBTS.

\paragraph{3G Networks}
The main feature of the 3G networks is the change of the focus from voice to data. In this generation, the network core is separated into two parts: one based on the conventional circuit switch architecture to support voice and one based on the packet switch architecture (best-suited) to support data. As a result, the 3G networks support both voice and data. Moreover, 3G realized ITU's IMT-2000 vision for cellular networks and supports new services such as mobile Internet through increased data rate. The dominant 3G standard is UMTS (Universal Mobile Telecommunications Service), which then evolved to HSPA (High-Speed Packet Access or 3.5G) as a transient generation with higher data rates. 3G networks are also still alive, again due to the vast investment and supporting IoT applications. OpenBTS-UMTS \cite{OpenBTS} is an open-source project that implements parts of the 3G networks.  

\paragraph{4G Networks}
The main feature of the 4G networks is an all-IP structure. In other words, the network core is purely packet-switched, named EPC (Evolved Packet Core). As a result, from this generation forward, we can state that cellular networks are the extension of the Internet and data services to mobile users. 4G networks satisfy the requirements of ITU's IMT-Advanced, and for multimedia communications such as voice, they rely on the IMS (IP Multimedia Subsystem) framework. Moreover, 4G networks demonstrate many advances in the communication theory in practice, such as small calls, relays, Carrier Aggregation (CA), Coordinated Multi-Point (CoMP), Inter-Cell Interference Coordination (ICIC) in the RAN, Control and User Plane Separation (CUPS) (as the first traits of SDN), and Dedicated Core Networks Selection (DECOR) (as the first traits of \textit{slicing}) in the core of the network. The dominant standard of 4G networks is LTE (Long Term Evolution), and the open-source 4G projects include OAI \cite{OAI4G} and srsLTE \cite{srsLTE4G}.

 \begin{figure*}[t]
	\centering{\includegraphics[width=0.8\textwidth]{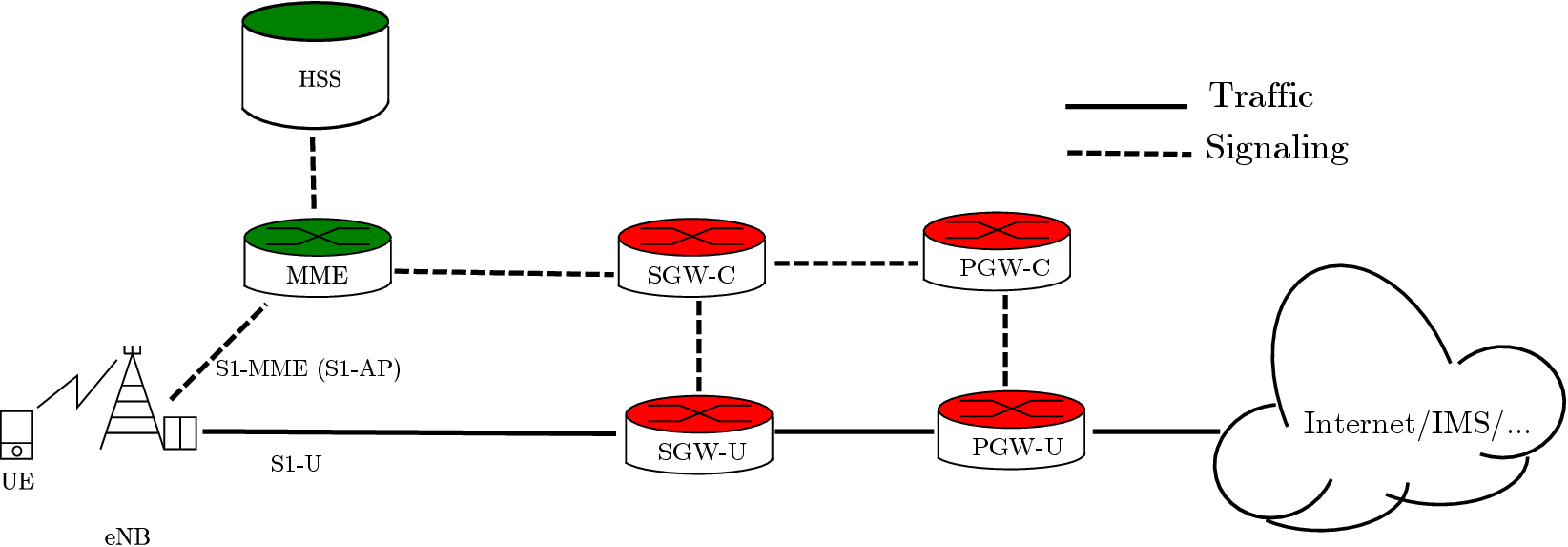}}
	\caption{The LTE's  CUPS architecture \label{fig:cups}}
\end{figure*}
As this paper implements 4G cellular VNFs on the \nm{}, we provide more details on EPC components here. Fig.~\ref{fig:lte} depicts a simplified LTE architecture. All links are logical and may be realized through an IP network. The PGW (Packet Gateway) is the gateway of the cellular network. Its main functionality is to be the mobility anchor point, which hides the mobility of the users from the outside world. PGW uses tunneling (encapsulation of the packets) to route the users' traffic to where they are located. Other functionalities of the PGW include QoS enforcement and IP address allocation to connected UEs (User Equipment). However, there may be millions of users, and supporting all of them with one or a handful of PGWs is not practical. This is where the need for SGWs (Serving Gateways) becomes evident. In fact, the whole area under the coverage of a cellular network operator is divided into regions, consisting of several cells. Each SGW is responsible for tracking the location of the UEs in a specific region and making necessary tunnels to their corresponding cell site base station (or eNB (evolved Node B)). Then, the end-to-end tunneling between the PGW and a user (UE) in a specific region is broken into two tunnels: the tunnel from the PGW and an SGW (the one responsible for that region) and the tunnel between that SGW and the eNB to which the UE is connected. As a result, the PGW needs to change its state to about one UE only when it crosses the region's borders. The component that is informed about the UE location by the eNBs and instructs SGWs and PGW(s) to configure their tunneling parameters is the MME (Mobility Management Entity). MME is also involved in paging, handover, authentication, security, and management of subscription profiles. Finally, since access to the network is not for free, we need a database that records subscribed users' information, including their identities, imprecise locations in the network, security features (secret keys), and QoS contexts. The HSS (Home Subscriber Server) realizes this database and its secure connection to the MME in the LTE architecture. 

The 3GPP release 14 introduced the idea of CUPS, the separation of the user plane and control plane as depicted in Fig.~\ref{fig:cups}. In CUPS, SGW and PGW are decomposed into two parts: one that routes users' data packets (SGW-U and PGW-U) and one that performs controlling the connections (SGW-C and PGW-C). Therefore, MME (as a pure controlling entity) establishes the connection to SGW-C and PGW-C, which control SGW-U and PGW-U, respectively. CUPS adds flexibility to network deployment and operation (as different components can be independently scaled up/down on demand and in different places), and it can be considered the first introduction of the SDN in cellular networks, which is matured/completed in the 5G networks (3GPP release 15 and above).

\paragraph{5G Networks}
The cellular networks take an ambitious step in 5G with the goal of supporting applications with different demands (first envisioned in ITU's IMT-2020). The realm of these applications is specified by the eMBB, the mMTC (or massive IoT (mIoT)), and the URLLC as the extreme corners where well-known applications (e.g., smart homes, augmented reality, and industry 4.0) can be placed somewhere between, based on their demands similarity to the mentioned corners (e.g., required peak data rate, connection density, mobility, and latency). After a period of ambiguity and discussion on ``what will 5G be" \cite{andrews}, it became clear that in order to support such diverse applications, SDN and NFV would be the key technologies in any true 5G realization. The SDN and NFV enable the creation of several logical networks, called network slices, on top of the same infrastructure, each tailored according to the requirements of a specific application. Additionally, in the RAN, 5G introduced New Radio (NR) which features even higher data rates, more flexibility (such as different spectral modes of operations and the alternative ways to interact with the core network; SA/NSA (Standalone/Non-Standalone)), massive MIMO, mmWave, and dual-connectivity (simultaneous connection of 5G UEs to a 4G eNB and a 5G gNB).

5G core network, also known as SBA (Service Based Architecture), consists of entities that less or more resemble their EPC's counterparts, such as Unified Data Management (UDM) ($\approx$ HSS), User Plane Function (UPF) ($\approx$ SGW-U and PGW-U), Session Management Function (SMF) ($\approx$ SGW-C and PGW-C), and Access and Mobility Management Function (AMF) ($\approx$ MME). It also comprises new entities related to virtual network functions and network slicing, such as Network Repository Function (NRF) (a repository listing all the functions), Network Slice Selection Function (NSSF) (which selects the network slice instances for a given UE), Network Exposure Function (NEF) (which exposes some internal events related to UEs), and Network Data Analytics Function (NWDAF) (which collects and analyzes information from the network and its users). The main characteristic of the SBA is that different entities interact with each other through well-established (and successful) API standards, such as HTTP and JSON, which makes the core of the 5G cellular networks more look like the usual client-server model in the Internet.

The 5G paradigm shifts towards more softwarization and virtualization, which has also revolutionized open-source activities. Projects such as OAI \cite{OAI5G} and srsLTE \cite{srsLTE5G} also have products for the 5G networks. Other projects such as free5GC \cite{free5GC} and Open5GS \cite{Open5GS} are also contributing to this subject (see \cite{bonati} for a thorough survey and comparison among different 5G open-source projects). 

Different open-source projects for cellular networks differ in community support, components they have implemented, and stability level. We found only one project claimed to be ``running for hours and days without any restart," which is the LTE's EPC (release 14) with MAGMA MME-based deployment \cite{refoai}. Hence, in our work, we have used this project for deploying on the \nm.

\subsubsection{RAN Splitting}
RAN splitting refers to the split of the RAN protocol stack, including the physical and MAC (Media Access Control) layers, into two or more parts that can be deployed separately within distinct nodes, and they can interact with each other over well-defined APIs. The idea behind the RAN splitting is to reduce capital expenditures and operating expenses (as a result of minimizing the cell site equipment), increase resource sharing and cope with the tidal effect (as a result of centralization), and enable advanced collaborative signal processing techniques, such as interference management and CoMP transmission and reception.

\begin{figure}[t]
	\centering{\includegraphics[width=0.48\textwidth]{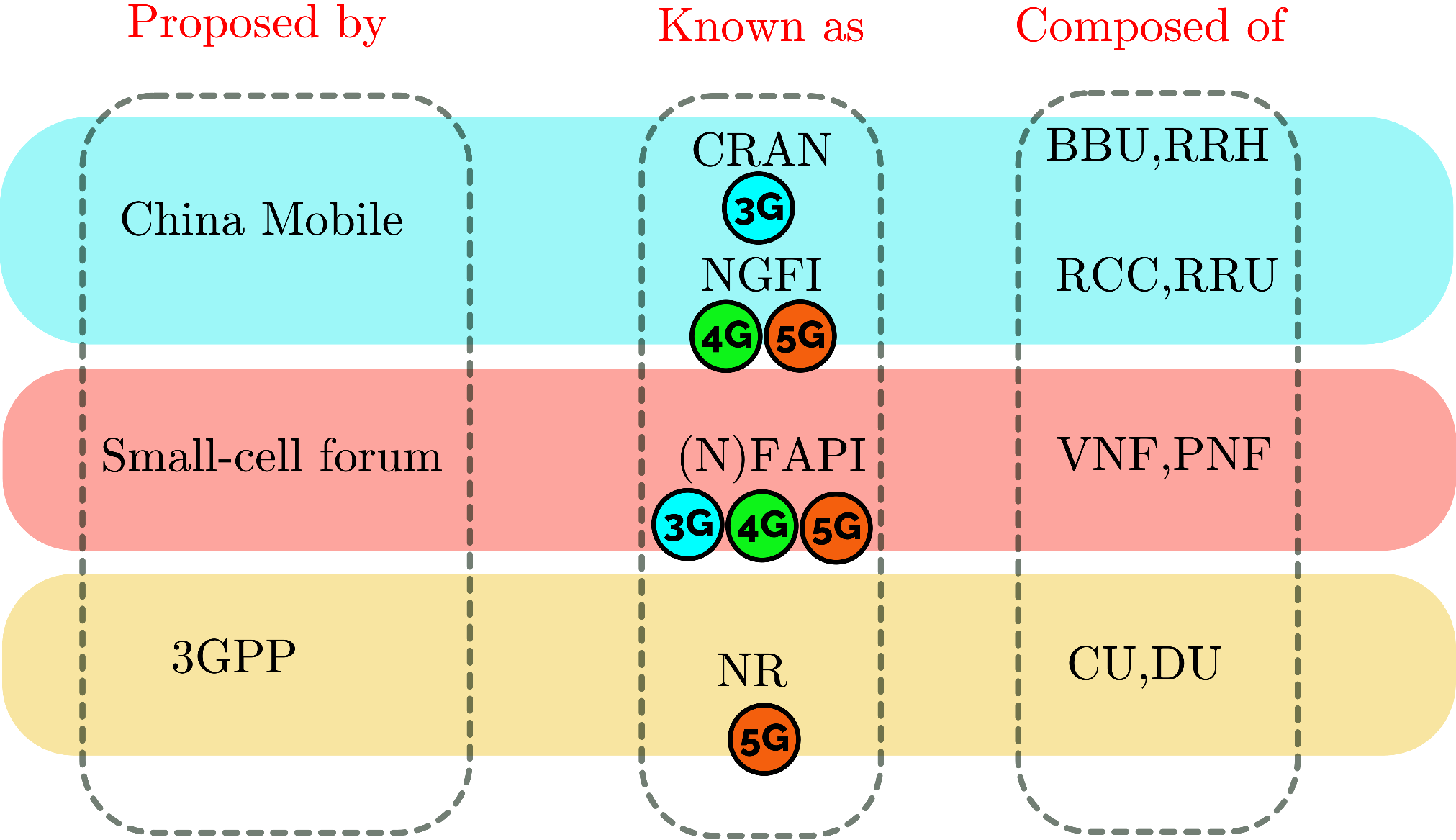}}
	\caption{RAN splitting proposals\label{fig:split}}
\end{figure}

For RAN splitting, different schemes have been proposed by different organizations, as shown in Fig.~\ref{fig:split}. China Mobile is the pioneer by first introducing the idea of RAN splitting and proposing the Cloud/Central RAN (C-RAN) for the 3G networks and, later, the Next Generation Fronthaul Interface (NGFI) for 4G/5G networks. In C-RAN terminology, the whole protocol stack breaks down into two parts named BBU (Base-Band Unit) and RRH (Remote Radio Head). The I/Q samples are sent back and forth between BBU and RRH using Common Public Radio Interface (CPRI) links. However, high stress on the fronthaul link (in terms of bandwidth and delay) resulted in limited deployments of C-RAN in practice. To cope with further bandwidth and latency constraints in 4G/5G, China Mobile redesigned the splitting in the NGFI scheme in which the separated components of the RAN are called the Remote Cloud Center (RCC) and Remote Radio Unit (RRU). China Mobile proposes six options in NGFI for the RAN splitting and compares the bandwidth and delay requirements of each option in its white paper. Specifically, the OAI project supports the following node functionalities:
\begin{itemize}
	\item  eNodeB
	\item  RCC and RRU (NGFI IF5): split-point at the OFDM symbol generator (i.e., frequency-domain signals)
	\item  RCC and RRU (NGFI IF4p5): time-domain fronthaul (more than 1 GbE is required)
\end{itemize}

On the other hand, Small Cell Forum (SCF) has been contributing by offering (Network) Function Application Platform Interface ((N)FAPI) which defines the API between the physical and the MAC (and above) layers for 3G, 4G, and 5G networks. Here, the separated parts of the RAN are denoted by Virtual Network Function (VNF) and Physical Network Function (PNF), respectively. 

Finally, at the time of introducing the New Radio (NR), 3GPP suggested ten options and suboptions for the splitting point for the 5G-RAN, which referred to separated parts as the Centralized Unit (CU), Distributed Unit (DU), and Radio Unit (RU). The O-RAN Alliance\footnote{The Open RAN (O-RAN) Alliance fostered efficient interoperability among multiple vendors by implementing standardized open interfaces in the 5G RAN. The O-RAN architecture further incorporates non-real-time radio intelligence controller (non-RT RIC) and near-real-time radio intelligence controller (near-RT RIC) modules, which are instrumental in facilitating sophisticated radio management for 5G and future 6G wireless networks.} has thoroughly assessed various RU/DU split configurations as suggested by the 3GPP. The focus was particularly on the alternatives for the physical layer division between the RU and the DU. The selection of the 7.2x split option results from its ability to harmonize the simplicity of the architecture of RU  with the requisite data rates and latency parameters on the RU-DU interface. On the other hand, for the split between the CU and the DU, the 3GPP option 2 is adopted by the O-RAN, whose interface is standardized as the \textit{F1 interface} \cite{polese2023understanding}. 


As mentioned, from the fifth generation onwards, cellular networks experience a transition from traditional hardware-centric approaches to flexible and dynamically configurable software-defined networks. In the following, we describe cloud technologies allowing a shift towards network softwarization and virtualization. Then, we explain how to use underlying technologies for cloud-native applications and incorporate them with cellular network projects to realize an end-to-end testbed for evaluating real-world network scenarios.

\subsection{Cloud-Native Applications and Infrastructure}
In this part, we look over the concepts and technologies that play essential roles in developing cloud-native applications and building the supporting infrastructure. 

\begin{figure}
	\includegraphics[width=0.48\textwidth]{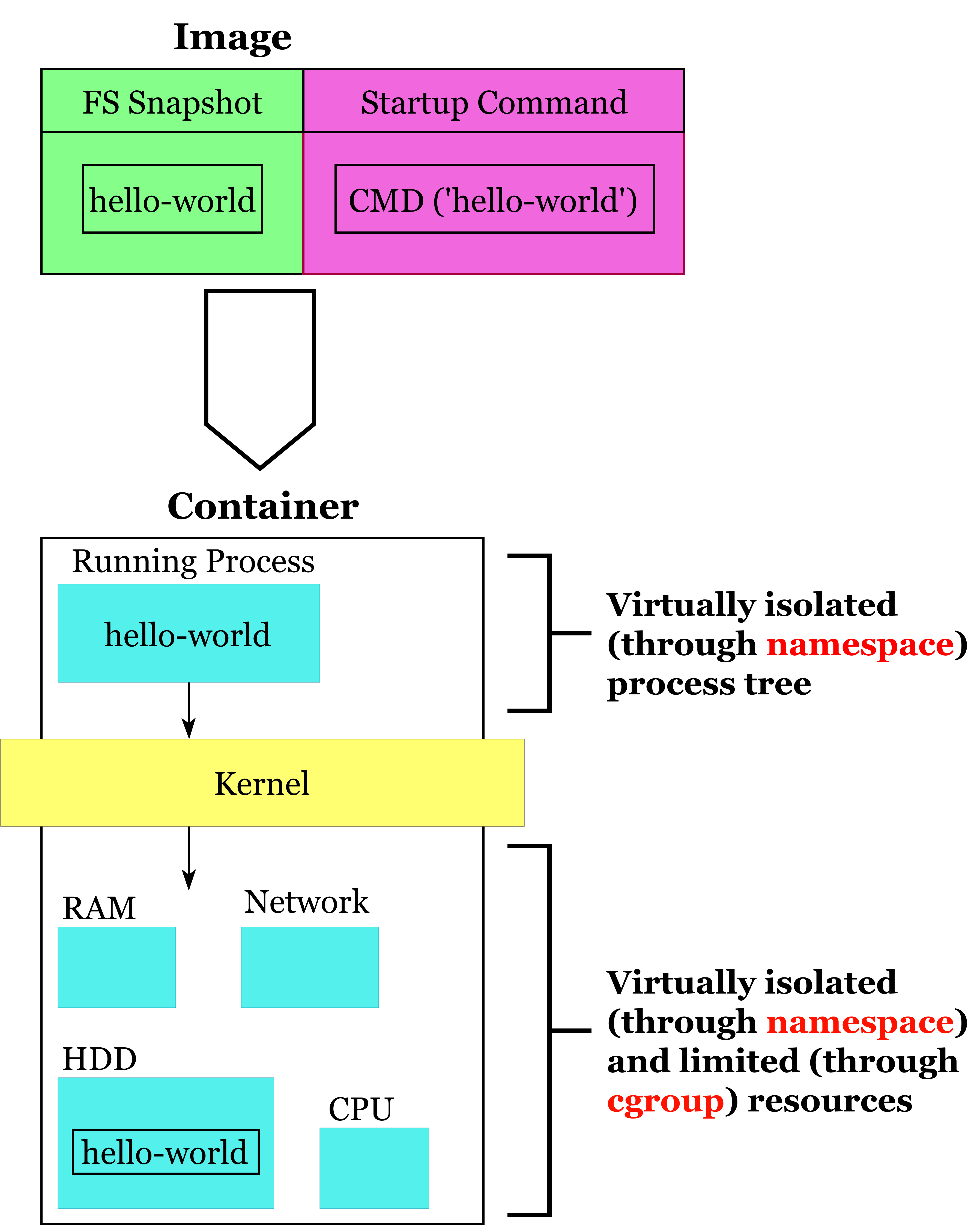}
	\caption{Image and container}
	\label{fig:fig3}
\end{figure}

\subsubsection{Container and Docker}
A container is a complete package of an application software itself and all its dependencies which can run on different computing systems quickly and reliably without complaining about necessary dependencies. The Linux primitives enabling containerization are \textit{namespaces} that provide isolation in different levels (e.g., process, filesystem, and network) and \textit{cgroups} that can be used to restrict resource consumption (e.g., memory, CPU, and bandwidth). 

Docker \cite{docker} is an open-source virtualization technology that facilitates the deployment, creation, and management of containerized applications. Similar to processes that are running instances of programs (executable files), the containers are also running instances of Docker images in their (possibly) isolated and restricted environments. An image is a compressed archive file containing application (and dependencies) files inside some directories along with a startup command that specifies the program by which the container's life begins executing (see Fig.~\ref{fig:fig3}). An image is created based on a recipe known as Dockerfile, which specifies the base image\remark{Base image and recipe needs refactoring}, files to be copied inside the image, instructions to build the application, and the startup command. Moreover, run-time parameters (e.g., networking, environment variables, and volumes) can be specified declaratively in a configuration \texttt{docker-compose.yml} file, which is fed to a tool called Docker Compose to run/manage containers accordingly. Docker also provides a registry (Docker Hub) to share (push/pull) official images built by different development teams.

\subsubsection{Kubernetes}
Kubernetes \cite{kuber} is an open-source platform for container orchestration, introduced by Google in 2014, to manage and automate the deployment, scheduling, monitoring, maintenance, self-healing, rollout and rollback, and operation of application containers across a cluster of machines. In fact, Docker enables developers and operators to create and run containers, while Kubernetes is used to orchestrate different containers. More precisely, Kubernetes allows the deployment of microservices over a number of machines and simultaneously hides the infrastructure from the application's point of view. Users can also deploy applications on different cloud providers using a standard set of APIs provided by Kubernetes, without separately needing to use each provider's API for deployment and management of the applications. 

In order to use the application management capability in Kubernetes, a description of the application's design is required. Then, Kubernetes turns the aforementioned description into a running set of objects and ensures they keep running by restarting those that fail. If some changes occur in the application's design, Kubernetes takes the required steps to transform (update) the set of running objects into new ones.
\begin{figure}[t]
    \centering
    \includegraphics[width=0.48\textwidth]{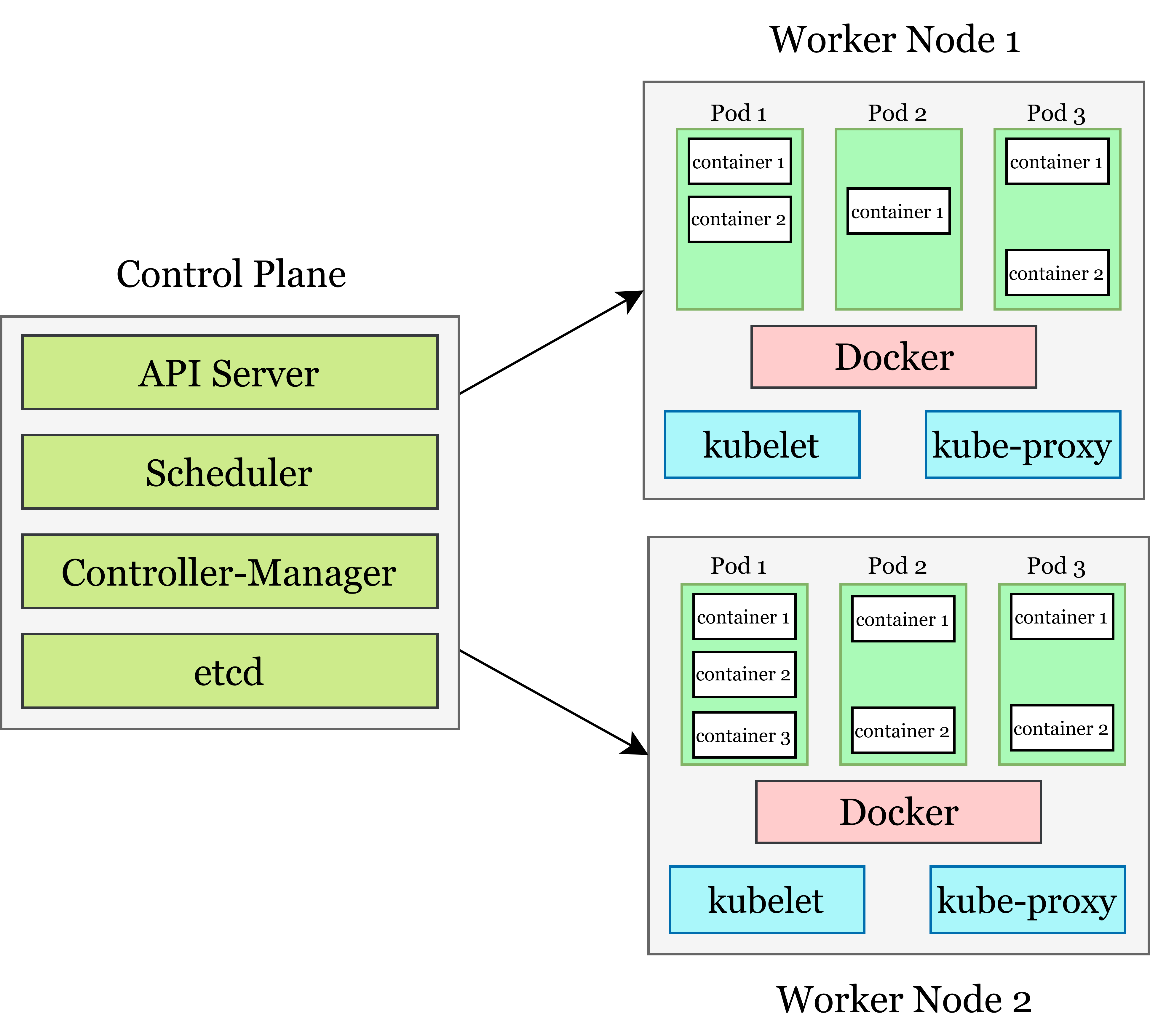}
    \caption{Structure of a Kubernetes cluster with two worker nodes}
    \label{fig:fig4}
\end{figure}

To distribute microservices among several machines, we need to create a Kubernetes cluster. The structure of a Kubernetes cluster mainly includes a number of worker nodes and a master node managing the application running over the workers. This structure is illustrated in Fig.~\ref{fig:fig4}. As shown in this figure, the master node includes several components, namely etcd (which is a key-value database recording the desired/current states), API server (which provides communication among etcd and other components\remark{Not technically accurate, maybe front-end to control plane is better!}), scheduler (which is responsible to place \textit{pods} and other objects on worker nodes), and controller (which is the brain of Kubernetes). A pod is a collection of one or more containers with some managing metadata, which represents the most basic deployable unit that can be created and managed in Kubernetes. In addition, a worker node consists of components called kubelet (the Kubernetes agent on each node, which is responsible for executing master's commands, such as pod creation/deletion using tools (e.g., Docker)) and kube-proxy (responsible for the networking and services). 

\begin{figure}[t]
    \centering
    \includegraphics[width=0.48\textwidth]{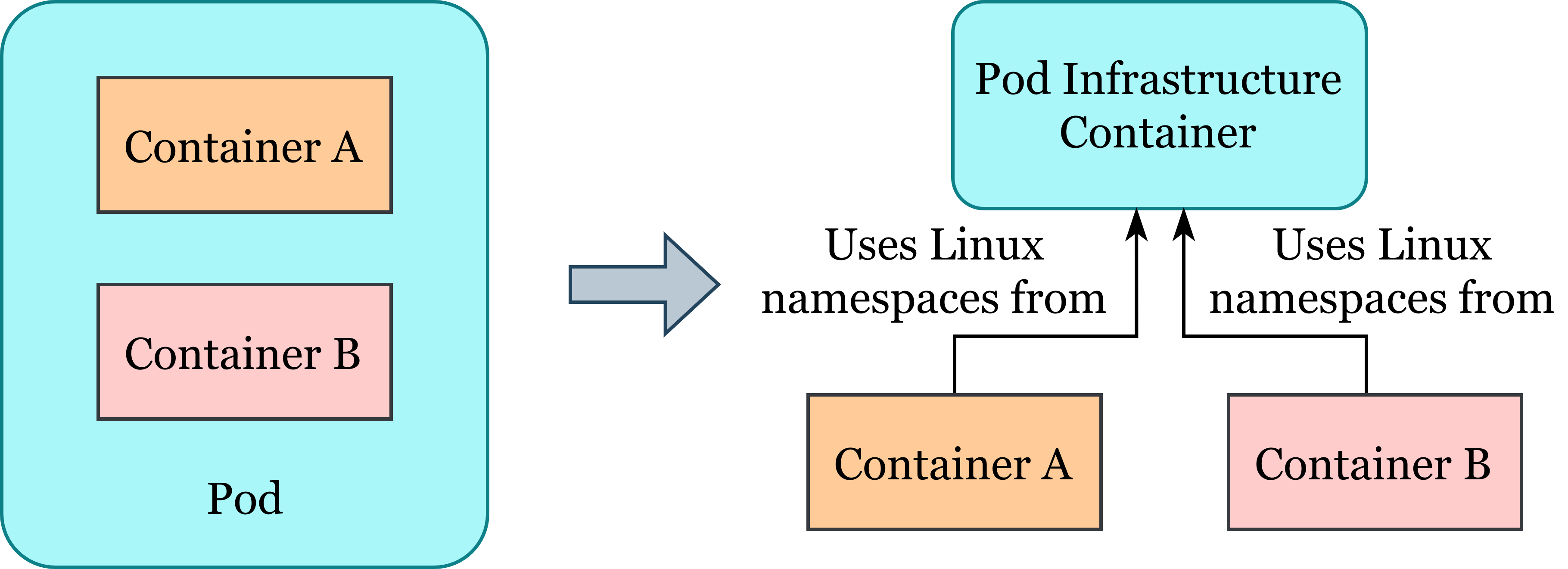}
    \caption{Structure of a pod containing two containers}
    \label{fig:fig5}
\end{figure}

Kubernetes has also spawned many other related open-source projects, mainly under the umbrella of the Cloud-Native Computing Foundation (CNCF), such as CoreDNS, Envoy, Helm, and Prometheus. Moreover, CNCF organizes several KubeCon $ + $ CloudNativeCon conferences per year.

\begin{figure}[t]
	\centering{\includegraphics[width=0.45\textwidth]{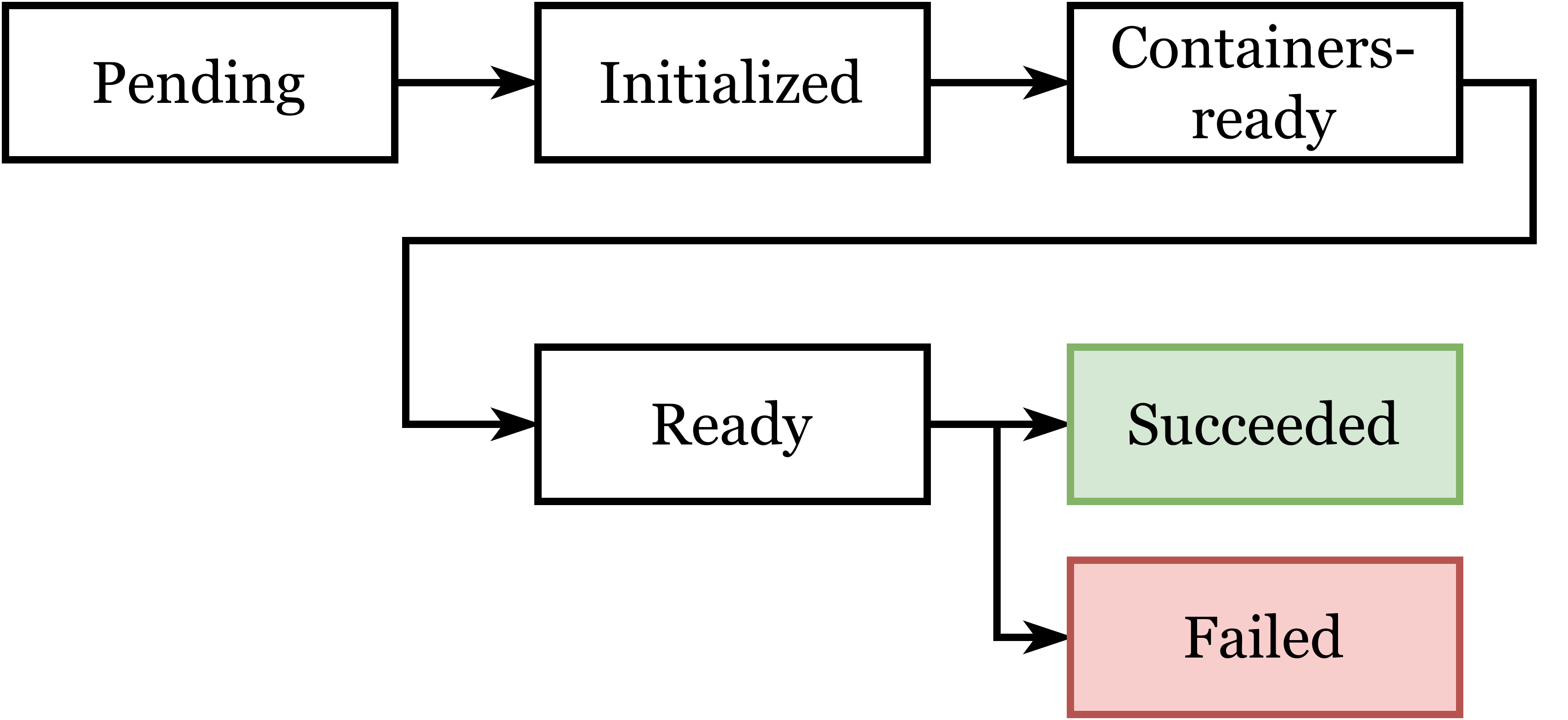}}
	\caption{Lifecycle of a pod \label{fig:fig6}}
 \vspace{-1em}
\end{figure}

\paragraph{Pod}
A pod is a group of one or more containers (with shared hostname, IPC (Inter-Process Communication), and network namespaces\footnote{Filesystem namespaces of pod's containers are different but they may have access to shared volumes.}) in addition to a set of specifications, including parameters such as labels and ports, among others. A pod can be viewed as an application-specific logical host and is the smallest deployable unit of computing in Kubernetes. The pod creation needs no more technology than what is required for the creation of containers, as it is indeed a main container (called the \textit{infrastructure} or \textit{pause} container) which hosts the application container(s) (see Fig.~\ref{fig:fig5}). Since Kubernetes is a declarative platform, to create pods, we instruct controllers, such as \textit{ReplicaSet}, \textit{Deployment}, \textit{Job}/\textit{CronJob}, \textit{DaemonSet}, and \textit{StatefulSet} through a \textit{pod template} in their YAML files to create and manage those pods on our behalf\footnote{Different controllers provide different functionalities, such as maintaining the number of pods, creating a specific pod on each node, and managing the pods during an application update.}. Fig.~\ref{fig:fig6} shows a typical pod's lifecycle. Before scheduling, it is in the \textit{Pending} (or \textit{Pre-scheduled}) state. Then, at the time of the creation of the pause container and execution of the \textit{init} containers, the state is \textit{Initialized}. Then, the main containers become ready at the \textit{Containers-Ready} state. After that, the pod is ready to serve at the \textit{Ready} state. In the case that all containers inside the pod terminate successfully, the state changes to \textit{Succeed}, and otherwise, to \textit{Failed}. Using the init containers which run before the main ones, we can make sure that all prerequisites are satisfied before starting the main containers. One usage of this feature in the pod's lifecycle is to enforce an order for the execution of the pods.

\paragraph{Networking}
Kubernetes's model for pod networking is flat, and all pods (logical hosts) are in the same subnet (i.e., connected through a \textit{logical} L2 switch). Pods inside a single node are connected to the same bridge through virtual Ethernet interface pairs. However, for pods on different nodes to communicate,  we need to somehow connect the bridges on different nodes to each other. As a result, providing Kubernetes network abstraction for pods can be complex and is usually achieved through an additional SDN layer on top of the actual network (which uses encapsulation and Linux networking tools such as iptables, routes, and IP-forwarding). Container Network Interface (CNI) is a project to simplify networking configurations in Kubernetes through the addition of plugins such as Calico \cite{calico}, Flannel \cite{flannel}, and Weave Net \cite{weave}. 

\paragraph{Volume and ConfigMap} 
By default, a pod's filesystem is the one defined in its image, and any file operation (e.g., creating a file or writing into a file) is ephemeral and will be lost by pod termination/restart. In Kubernetes, we can use different types of \textit{volume} to have different levels of file persistency. For example, \textit{emptyDir} volume provides persistency across container restarts during the pod's lifetime. Moreover, \textit{hostPath} and \textit{nfs} volumes provide higher persistency across pod restarts in a node and across node change, respectively. Similar to networking, storage consistency is complex in general, and several Container Storage Interface (CSI) projects, such as Cinder \cite{cinder} and Cephfs \cite{ceph}, have been developed to accomplish the work as plugins.

A related topic to volumes is the \textit{ConfigMap}, which is used to pass arguments to containers by mounting configuration files into containers through a special type of volume, not surprisingly, called ConfigMap.

\subsubsection{Mininet}
In SDN-enabled networks, the control plane handling management operations is logically centralized and physically isolated from the data plane, which in turn enables high network configurability and programmability. The network programmability and the capability of optimizing resource allocation and utilization in a centralized way, made possible by the SDN paradigm, are expected to alleviate the burden of the data onslaught expected from data-intensive applications \cite{ref30}. Mininet \cite{mininet} is an emulation tool that allows one to virtually emulate a complete SDN network scenario comprising a number of virtual hosts, controllers, switches, and links. It uses container-based virtualization to make a single system act as a complete network. 

\begin{figure}[t]
 	\centering{\includegraphics[width=0.39\textwidth]{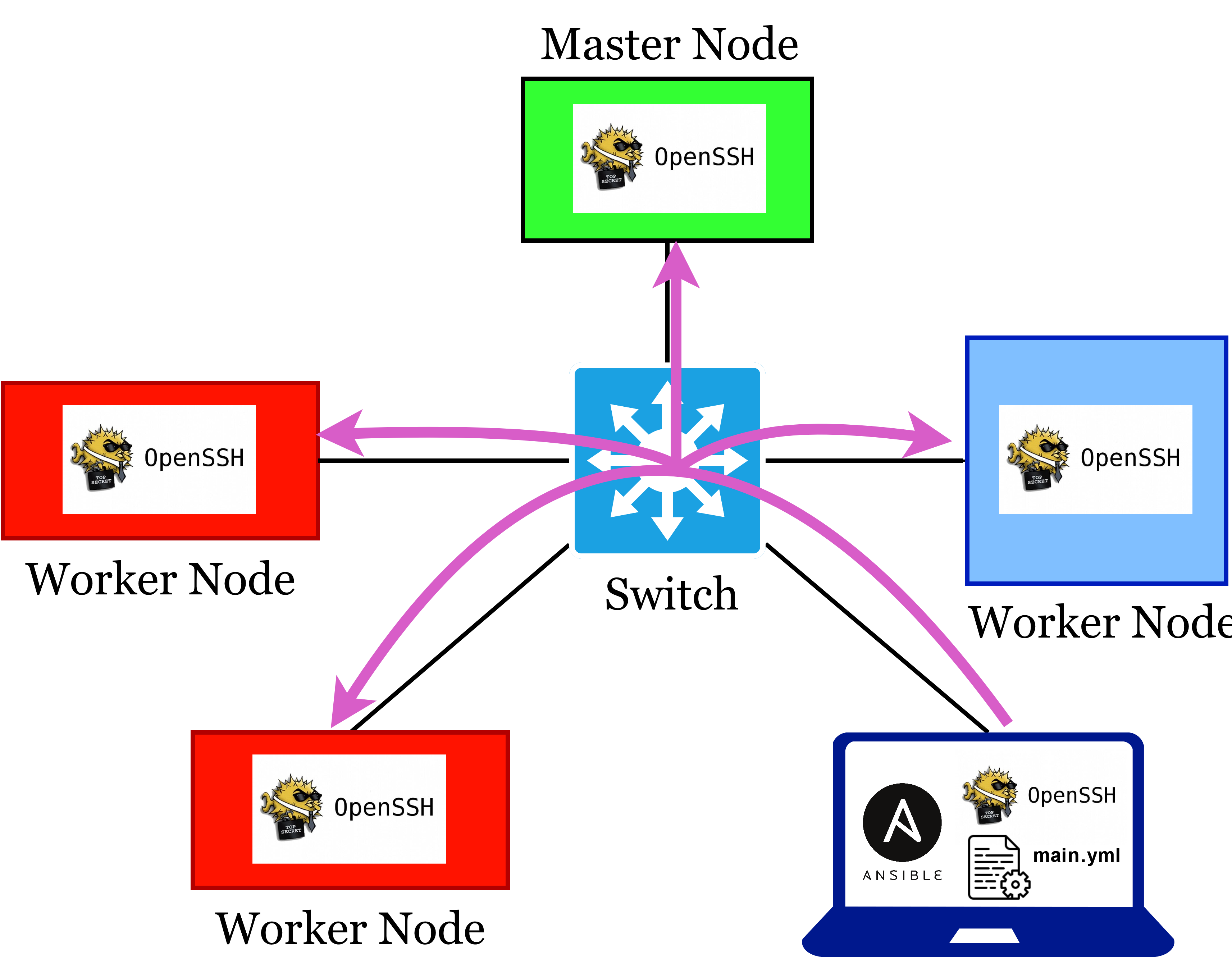}}
    \tiny
    \vspace{-0.5em}
    \begin{verbatim}
    $ ansible-playbook -i remote-hosts main.yml --ask-become-pass --ask-pass
    \end{verbatim}
    \vspace{-2em}
 	\caption{Installing required packages and configuration on cluster nodes using Ansible\label{fig:ans}}
\end{figure} 

\section{\nm{} Overview}\label{Sec3}
The structure of \nm{} aims to facilitate the building of flexible network scenarios with various network topologies. In this section, we describe the steps to construct the \nm{} alongside our best practices (i.e., the experimentally best-realized solution to cope with a challenge) in each step. In addition, we present a detailed tutorial demonstrating how to get from a portable network implemented with Docker containers to a cloud-native network implemented on the Kubernetes platform. The proposed structure of \nm{} resembles the \textit{flexible} and \textit{hierarchical} model of the modern cellular network (Fig.~\ref{fig:fig1}).

Specifically, \nm{} has been designed to demonstrate a network with edge computing, cloud computing, RAN splitting, and NFV capabilities. Two bridges (each emulating an IP network) are incorporated in the \nm{} structure; FH and TN. The FH bridge sits between distributed RAN components, such as RCC (or CU) and RRU (or DU), and therefore, it enables the RAN splitting test cases. On the other hand, the TN bridge may divide the structure into two parts; one close to RAN (or UEs) and one far away. As a result, the distinction between edge computing and cloud computing can be investigated.
  
Besides the bridges, the \nm{} includes a cluster of four machines. The cluster size is minimal as we need one machine for the master node, two machines (i.e., two worker nodes) to host different components of the distributed RAN, and one machine (i.e., another worker node) to host the VNFs (such as core network's VNFs), which may reside in the data center in reality. All nodes are connected to a Gigabit Ethernet switch. One node is further connected to the USRP-B210 (via USB 3.0) that plays the role of radio head in our setup.

In order to automate the process of installing required packages (e.g., Git, Python, Wireshark, and Docker), set configurations regarding CPU states and the Linux kernel, and copy some bash scripts (for enabling IP-forwarding, IP-addressing, etc.), we first list all of these prerequisites in an \textit{Ansible playbook} file in the \textit{Ansible management node} (in our case a laptop with the SSH connection to the cluster nodes, aware of their IP addresses, usernames, and passwords, which has Ansible and openssh-client installed). In this way, after installing the OS on bare-metal machines (cluster nodes), it suffices to install openssh-server on them. Then, the Ansible management node can perform all the required installations and configurations specified in the playbook on all cluster nodes using a single command as demonstrated in Fig.~\ref{fig:ans}.

\vspace{-1em}
\subsection{Challenges and Limitations}

1) Our main focus in this work is how to set up the \nm{} framework regardless of the particular VNFs that are used to realize the final prototype of a specific 4G/5G network. Since the 4G/5G open-source projects are evolving and have different stability levels, we resort to using OAI's (MAGMA MME-based) 4G implementation as it is claimed to be ``running for hours and days without any restart" \cite{refoai}. Moreover, the 4G core network is still used in 5G Non-Standalone (NSA) systems. 
 However, \nm{} is designed as an extendable framework. One can replace VNFs with those implemented for the next generations of cellular networks. We explain how to use the general structure of \nm{} with desired modules (VNFs) to deploy a specific generation of cellular networks in a cloud-native testbed.
\\ 
 \indent 2)  The number of UEs and eNBs in \nm{} is currently constrained by the availability and capabilities of commercially available hardware. Additionally, the state of the open-source project we are utilizing imposes further restrictions on scalability. In particular, we have tested \nm{}  with one eNB and two COTS UEs\footnote{In our setup, we use the USRP-B210 without any external/GPS clock and duplexer. Therefore, some COTS UEs face problems in detecting and connecting to our network. Among different UEs that we have tested, the Huawei-E5573-8739 and Huawei-Nova 3e can successfully connect to our network (both using Qualcomm Snapdragon 680 chipset)}. Additional eNBs and UEs may be added to  \nm{} as the hardware and/or software capability evolves.
  It is also worth noting that similar works in the field often face the same limitations regarding the number of UEs and eNBs due to these practical constraints \cite{10183111,ref22,ref4,ref16,8524891}.
 \\\indent
 3) One significant challenge in creating a testbed from open source projects is the versioning issue, as different components are continuously evolving. A testbed that functions correctly may break after an update to one part. To address this challenge and ensure the reproducibility of our testbed, we have consolidated all configuration commands and image versions into several files. These files are utilized by an Ansible playbook to automate the installation of the necessary packages, thereby maintaining version consistency across all components.

\subsection{Cluster Setup}
In this section, we describe the steps to set up a Kubernetes cluster. As explained in the background, a Kubernetes cluster comprises a set of components, such as kube-proxy (for managing network connectivity) and kubelet (for ensuring the overall health and stability of containers in pods), which should be installed and configured on each node. Also, in case of restarting a node, cluster components need to be recovered. Since manually installing these components is time-consuming and error-prone, several tools are provided for the cluster installation. We tested three tools, including Kubeadm \cite{Kubeadm}, Kubesphere \cite{KubeSphere}, and Rancher Kubernetes Engine (RKE) \cite{rke} to install clusters. Among them, RKE is selected as this tool is handier, more stable, and easier to configure. Furthermore, due to the flexibility of RKE, we can configure multiple cluster options in the rancher's configuration file which is used to deploy a cluster.

The Kubernetes networking model requires certain network features but at the same time allows a degree of flexibility regarding the implementation. As a result, various projects have been released to address specific requirements. Container Network Interface (CNI) is one of those projects supporting plugin-based functionality to simplify networking in Kubernetes. The main purpose behind CNI is to provide enough control to administrators for monitoring communication while reducing the overhead of manually generating network configurations. A main challenge when designing \nm{} was to pick up the best CNI for the framework. Among different CNI providers, Calico \cite{calico} is used in the final structure of \nm{} as it enables us to assign a static IP to each deployed pod, and this easy-to-deploy CNI provider is supported by RKE. 

We deploy a Kubernetes cluster using four machines with Intel Core i9-11900 CPU@3.5 GHz, 32 GB of RAM DDR4 memory, using an operating system Ubuntu 18.04 LTS, interconnected by an L2/L3 switch. This cluster consists of a master node, three worker nodes, and two bridges (FH and TN) as illustrated in Fig.~\ref{fig:fig1}.

\subsection{Building Docker Images}
Docker provides two ways to run a container using Docker images. One is to pull the Docker image of the application from a Docker registry (e.g., Docker Hub), and the other is to build the Docker image of the application using Dockerfile. A Dockerfile is defined as a recipe containing all the required dependencies and some instructions in order to make the application runnable inside the Docker container.

\begin{figure*}[t]
\centering{\includegraphics[width=\textwidth]{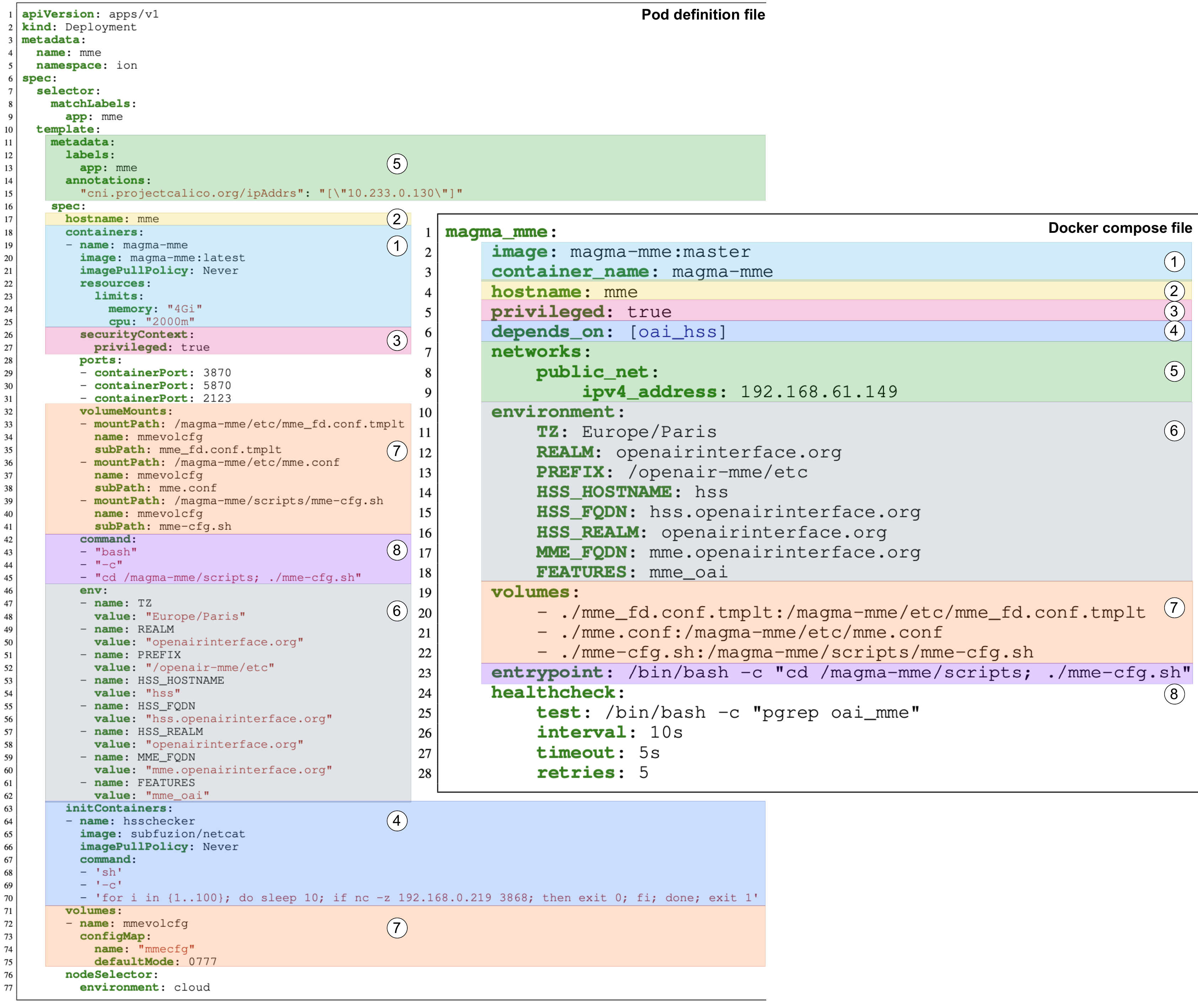}}
    \tiny
 	\caption{Docker compose file for MME module and its corresponding pod definition file\label{fig:tag}}
\end{figure*}

We used OAI's LTE Evolved Packet Core (OAI-EPC) as the virtual network functions in \nm{}. OAI has a set of public GitHub repositories, each of which provides the codebase of the implementation for a specific core module \cite{OAI4G}. OAI also provides Docker images for the core modules in the Docker Hub. Thus, for each module (e.g., MME, HSS, SPGWC, SPGWU, etc.), we directly pulled the corresponding Docker image from Docker Hub. 

On the other hand, to build the RAN module, we take Dockerfiles existing in EURECOM GitLab \cite{ref18}. Specifically, the final RAN image is built by a multi-stage Dockerfile build process, which brings about a lightweight final image for the RAN module and accelerates the process of image rebuilding. It is worth mentioning that the Dockerfile written to build the eNB image and those for RRU and RCC are being updated by EURECOM periodically. Thus, we tested different versions to find those working properly. This was one of our challenges during the \nm{} framework setup. We also pushed the final images for both CN and RAN modules to the \nm{} repository with new tags. Thus, it is no longer required to build images from scratch when setting up the framework.
 
\subsection{Moving from Containers to Pods}
As mentioned in the background section, a pod is a Kubernetes object that encapsulates one or multiple containers. The first step to deploy a module on Kubernetes is to create a pod definition based on the docker-compose file, which is then placed inside the \textit{pod template} part of a deployment object. All the needed configurations (e.g., environment variables and volumes) must move properly and precisely from the docker-compose file to the pod definition file. Here, we take the MME definition file as an example since MME is the most complex among OAI modules, and it covers all significant issues for converting docker-compose file to the pod definition file. Fig.~\ref{fig:tag} depicts how to write a pod definition based on a provided  docker-compose file.

A pod definition file consists of four sections; \texttt{apiVersion}, \texttt{kind}, \texttt{metadata}, and \texttt{spec}. The \texttt{spec} is the part under which an array of containers could be configured and is worth concentrating on. There exists a \texttt{nodeSelector} option, under the \texttt{spec} part, which determines a specific node where each pod is deployed (see line 76 in the MME definition file illustrated in Fig.~\ref{fig:tag}). This option is one of the basic parts of a pod definition file. In the following, different sections of the MME definition file in Fig.~\ref{fig:tag} and their corresponding in the docker-compose file are discussed.

\subsubsection{Networking}
In Docker, we could create a bridge network and assign a unique IP address to each container. However, assigning static IP addresses to pods in Kubernetes is challenging due to its architecture and purpose. On the other hand, OAI software modules must bind to a specific IP. To this end, we used Calico as a CNI so that each deployment object is assigned a unique IP address in the range of the Calico subnet (see \circled{5}).
 
\subsubsection{Security}
Similar to \texttt{privileged: true} in the docker-compose file, we define a \texttt{securityContext} in the pod definition file, which conducts Kubernetes to run the containers in privileged mode. This is required for containers to have some capabilities, such as system administrator (sysadmin), to operate as expected (see \circled{3}).

\subsubsection{Data and Volumes}
The \texttt{volume} section in the docker-compose file can be translated into the \texttt{hostPath} and \texttt{mountPath} sections. The \texttt{hostPath} is located on the node that the \texttt{nodeSelector} option refers to. Also, we used another type of volume named \texttt{ConfigMap}. There exist some configuration files that should be converted into \texttt{ConfigMap} in the Kubernetes cluster to be used as volume bindings in the modules (see \circled{7}). Note that to configure the RAN module, it is required to bind the \texttt{/dev/bus/usb} path to the eNB container as we use a USRP board connected to the USB 3.0 port. Hence, we use the \texttt{mountPath} field in the eNB definition file to bind the aforementioned path to the eNB container.

\begin{figure}[t]
    \centering{\includegraphics[width=0.48\textwidth]{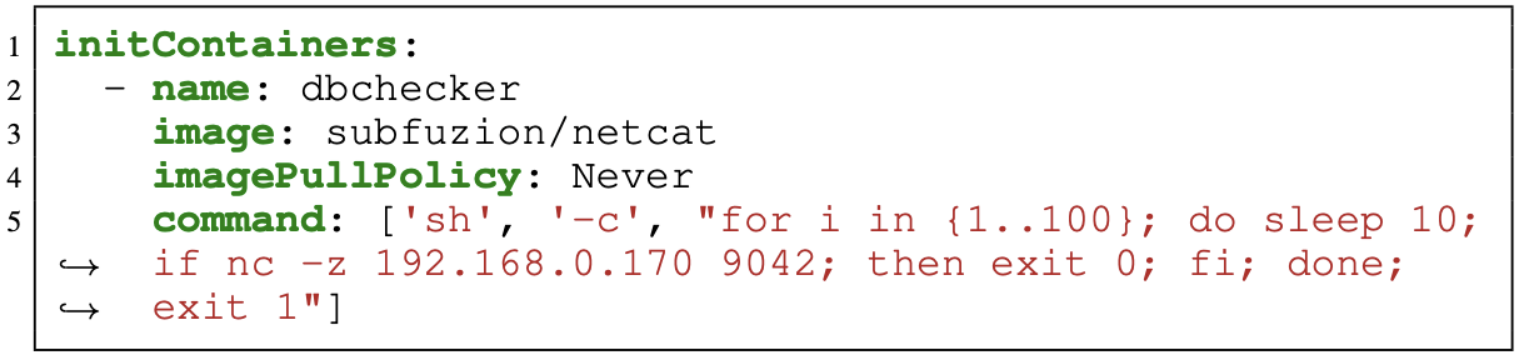}}
    \caption{Init container in HSS deployment file \label{fig:hssinit}}
\end{figure}
    
\subsubsection{Environment Variables}
In Docker, environment variables can be set using either the \texttt{environment} attribute or the \texttt{env\_file} option. This can be represented as an \texttt{env} section in the pod definition file, which allows to set environment variables for containers by specifying a value directly for each variable (see \circled{6}).

\subsubsection{Dependencies}
\begin{figure}[t]
	\centering{\includegraphics[width=0.48\textwidth]{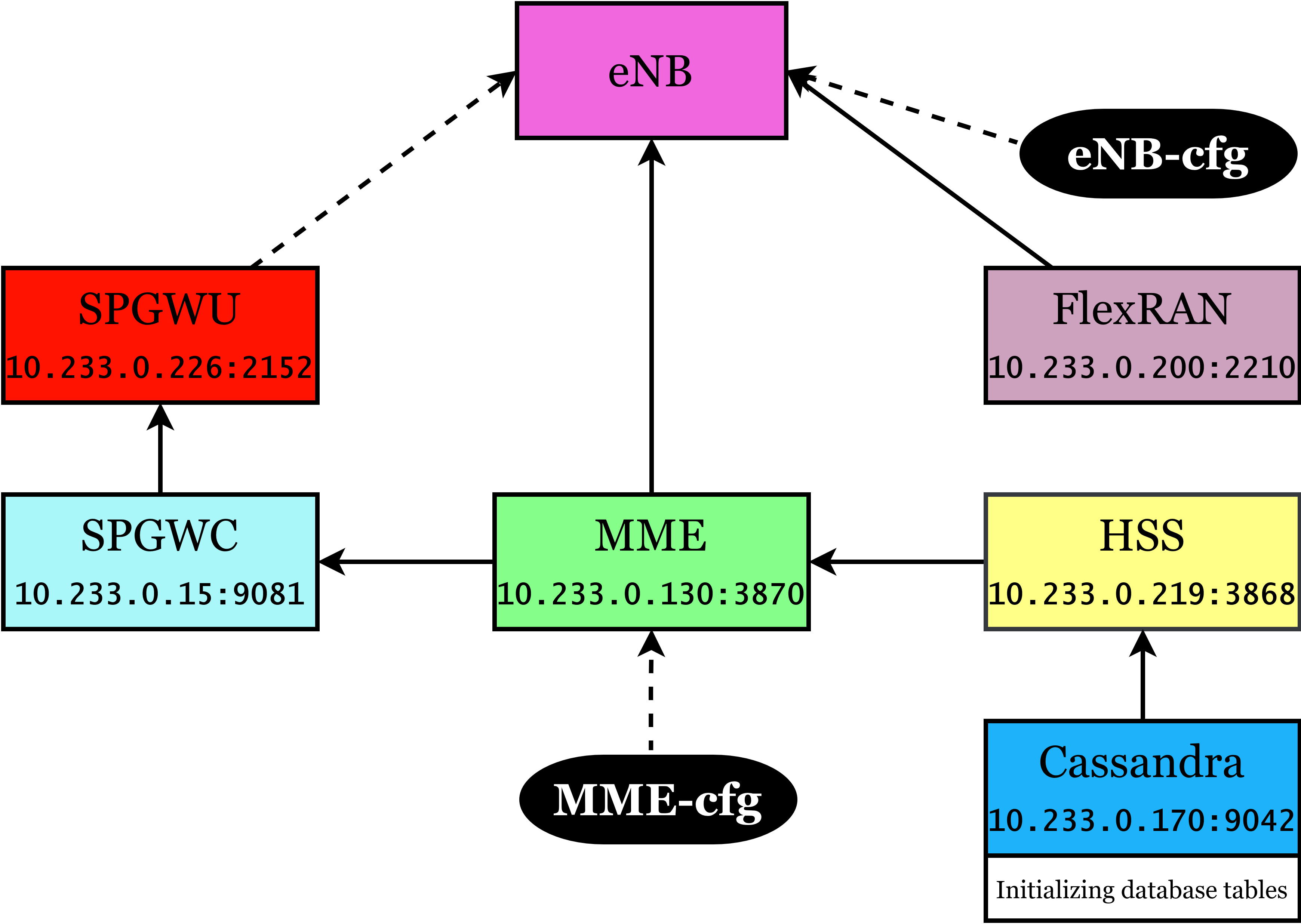}}
    \caption{Dependency tree of the VNFs \label{fig:init}}
\end{figure}

The \texttt{depends\_on} field in a docker-compose file sets the order of container deployment. However, in Kubernetes, we use \texttt{initContainers} to achieve the same functionality. Init containers check if a specific port of the container where they depend is open using the \texttt{netcat} command. Once the init container successfully returns, the pod will be deployed (see \circled{4} and dependency tree among different VNFs as demonstrated in Fig.~\ref{fig:init}).

\subsection{Deploying the Cellular Application}
After installing the cluster and preparing deployment YAML files, a few steps are required to deploy the whole application and check the connectivity of modules to each other. Also, we faced some challenges during the deployment phase, the most important of which are elaborated on in the following.

\subsubsection{Database and HSS} 
The HSS needs to connect to a database server to store and retrieve subscribers' profile data (e.g., IMSI, APN, and secret keys). In our setup, we used Cassandra as the database server and ran a procedure to create some specific empty tables for the HSS to access. In Docker deployment, a \textit{db\_init} container creates those tables. However, in Kubernetes, those (empty) tables could be created using an init container that executes the db\_init instructions. Therefore, in our setup, we first deploy the Cassandra with an init container. Then, the HSS module is deployed, which fills the aforementioned empty tables in the database. Fig.~\ref{fig:hssinit} shows the implementation of the HSS's dependency on Cassandra using the init container primitive.

\subsubsection{MME, SPGWC, and SPGWU} 
As depicted in Fig.~\ref{fig:init}, MME should be deployed with its ConfigMap after the deployment of the HSS. The HSS will log \texttt{STATE\_OPEN} once the MME is up and running. After the MME, the SPGWC and the SPGWU can be deployed, respectively. These modules send some \textit{heartbeat} packets to each other which can be observed on their container logs as a health check procedure. 

\subsubsection{FlexRAN} 
FlexRAN \cite{flexran} is a flexible and programmable platform that separates the RAN control and data planes and supports the design of real-time RAN control applications (such as defining multiple RAN slices with different Recourse Blocks (RBs) and assigning each UE to a specific slice). It listens on two different ports, one for the eNB agent and one for API requests which we use to interact with this module. There is a \texttt{NETWORK\_CONTROLLER} section in the eNB (or RCC) configuration file in which we can specify if we want to connect to a FlexRAN agent by setting the \texttt{FLEXRAN\_ENABLED} option to \texttt{yes} or \texttt{no}. It is worth noting that FlexRAN deployment does not depend on any module in our application, and therefore, it can be deployed at any stage before the eNB (or RCC).

\subsubsection{eNB} 
According to Fig.~\ref{fig:init}, the last step in the application setup is to run the eNB deployment. The eNB deployment refers to a ConfigMap object (the eNB-cfg in the Fig.~\ref{fig:init}) which defines all eNB parameters, such as MNC, MCC, and MME IP address, and therefore, it should be deployed first. The eNB VNF needs to be deployed on a system with low-latency Linux kernel and all CPU cores in the C0 state (checked with the i7z utility). On the other hand, the eNB module uses an \textit{image-downloader} script to download the FPGA image of the USRP. This program itself downloads the required UHD image based on the USRP type, which is set as an environment variable. In order to speed up the procedure, one can download the suitable image according to the board type which is referenced as a hostPath volume (placed in the worker node hosting the eNB).
 
In a scenario with distributed RAN which uses RCC and RRU, the last step is to run their deployment files. These modules significantly depend on the kernel and OS settings. Thus, we must consider items such as low-latency kernel and power management settings in the BIOS and Grub in their worker nodes. One of the challenges is to use the most compatible version of the low-latency kernel which is ``4.15.0-206'' based on our experience. To have a working RCC and RRU pair, they should run based on the same Docker image (using different configurations). Again, finding the best candidate among different image files is a crucial step and the result of our investigation is the one included in the \nm{} Docker Hub. 

\subsection{Mininet Bridges}
In this framework, we address the effect of the distance between different components in a real network by utilizing SDN tools and the Mininet emulator and creating virtual networks between our devices. We name these virtual networks Mininet Bridge (MB). In this part, we present how an MB is created and what it consists of.

\subsubsection{Mininet}
To create an MB, we require a tool capable of emulating real-world computer networks. Mininet is a lightweight network emulator able to set up virtual networks containing virtual hosts, switches, controllers, and links on an OS. Mininet takes advantage of Linux namespaces instead of virtualization, resulting in a lighter emulation compared to Virtual Machines (VMs). Mininet's straightforward Python API enables the creation of complex and real-world topologies. Moreover, Mininet has a Command Line Interface (CLI) by which users can manage, configure, and interact with the created virtual network. Mininet can also utilize an SDN controller to emulate SDN networks. In a virtual SDN network created by Mininet, the switches are virtual OpenFlow switches capable of communicating with controllers using the OpenFlow protocol. 

\subsubsection{SDN controllers}
There exist many SDN controllers with different capabilities (e.g., POX, Floodlight, and RYU) to serve as controllers connected to the virtual SDN network. In this framework, we used ONOS \cite{onos} and RYU \cite{ryu} as the SDN controller of the MBs. ONOS and RYU are both popular open-source SDN controllers that support the OpenFlow protocol and have a large community backed by the Linux Foundation. RYU is Python-based, making it easier to set up and use, which results in more popularity among developers. ONOS is written in Java and is more complex compared to RYU, but it has more features and can handle large-scale networks. Moreover, ONOS is a part of the Open Networking Foundation which is supported by many major vendors and leading companies in the telecommunication industry. It is worth mentioning that ONOS is more suitable for our framework as is more compatible with large-scale telecommunication. 

\subsubsection{Bridge}
To create an MB, we need a Linux OS with Mininet and ONOS/RYU installed on it. We create our topology with Mininet's Python API without hosts, and using Open vSwitch (OVS) commands, we connect the Network Interface Cards (NICs) of the device to the virtual switches. The use of MBs makes our framework more realistic and more similar to real-world communication networks. Moreover, MBs are considered transparent and are useful components in time of network management, measurement, and manipulation. 

\section{Evaluation}\label{Sec4}
In this section, we use the \nm{} framework to investigate the importance of VNF placement and RAN slicing as two cellular network capabilities. The details on how to deploy test cases along with the required scripts to reproduce the results are available in our GitHub repository.
\begin{figure*}
    \centering{\includegraphics[width=0.92\textwidth]{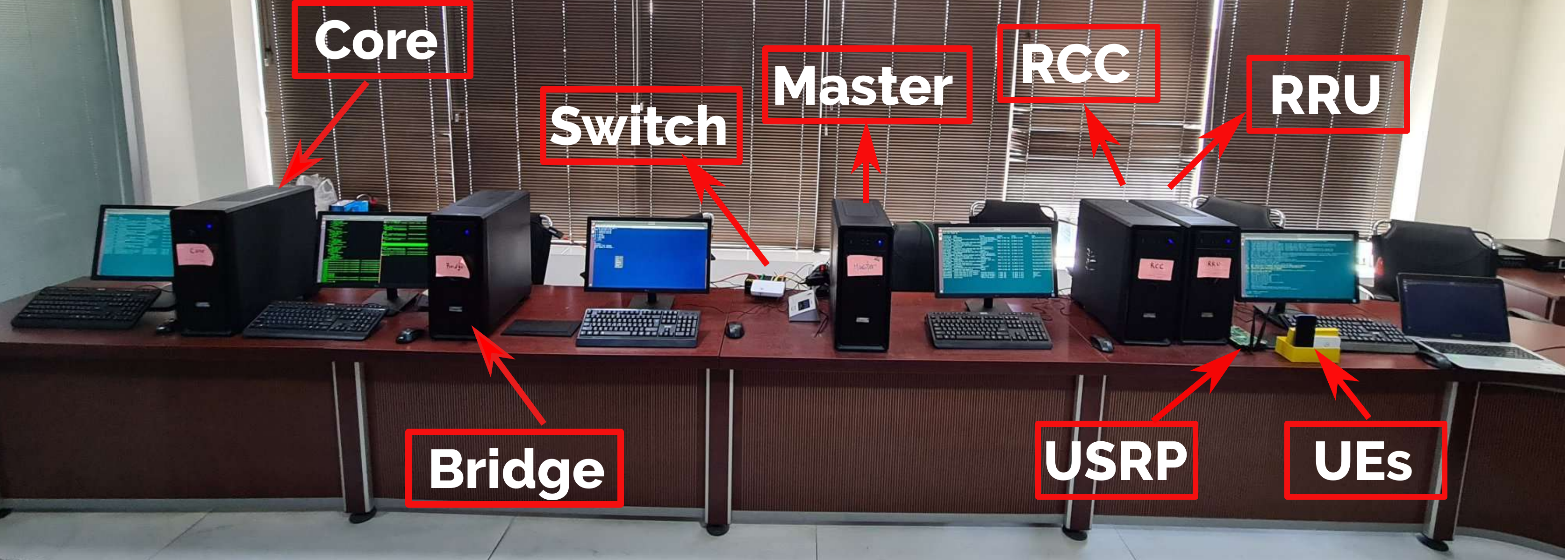}}
	\caption{\nm{} setup for the VNF placement evaluation \label{fig:demo}}
\end{figure*}
\begin{figure*}[t]
	\begin{subfigure}{0.48\textwidth}
		\centering{\includegraphics[width=\textwidth]{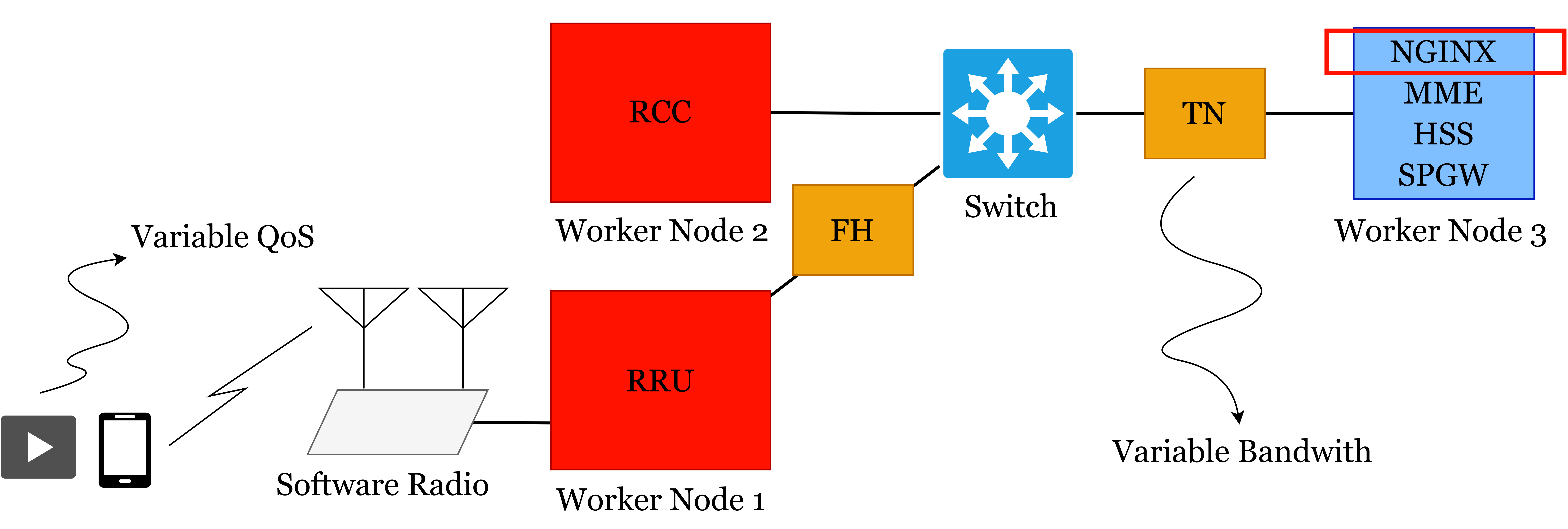}}
		\caption{Cloud computing\label{fig:demo2p1}}
	\end{subfigure}	
    \hfill
	\begin{subfigure}{0.48\textwidth}
		\centering{\includegraphics[width=\textwidth]{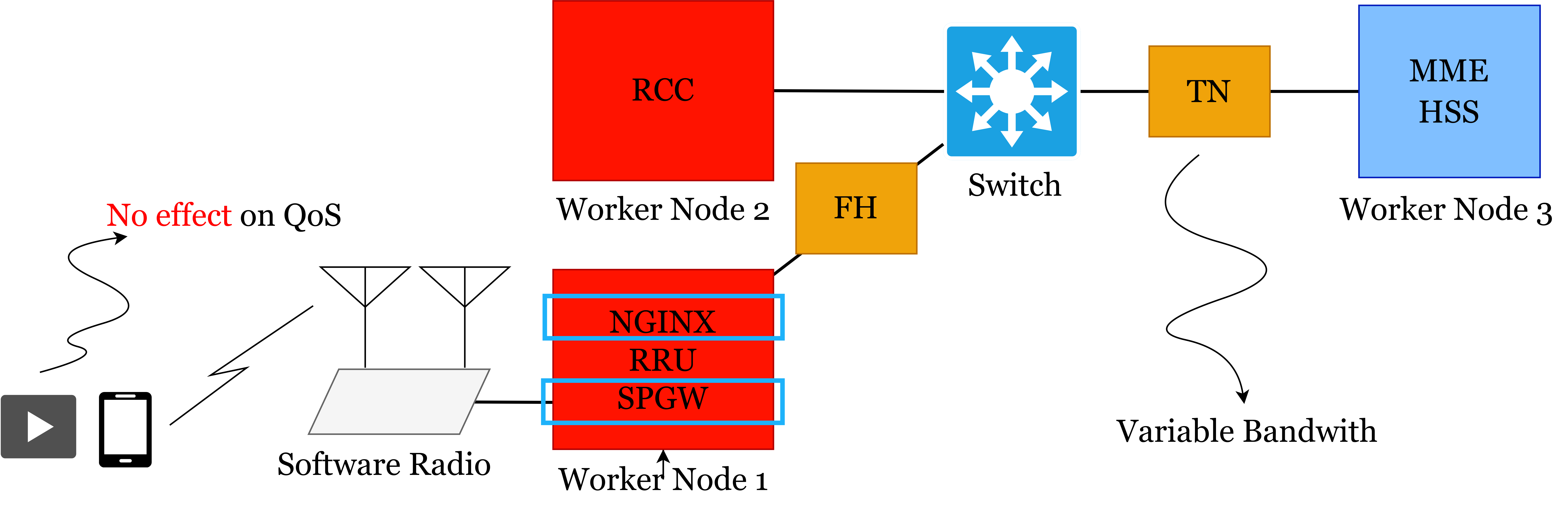}}
		\caption{Edge computing\label{fig:demo2p2}}
	\end{subfigure}
    \hfill
	\caption{Test cases for the VNF placement}
	\label{fig:demo1}
\end{figure*}

\subsection{VNF Placement}
Fig.~\ref{fig:demo} shows the \nm{} setup to demonstrate the impact of VNF placement. Specifically, the test cases incorporate two scenarios which are schematically represented in Fig.~\ref{fig:demo1}. In scenario 1 (Fig.~\ref{fig:demo2p1}), the multimedia server (an NGINX stream module) is placed on worker node 3 which also hosts the core network's components. On the other hand, the client is inside the UE. Therefore, the data path between the client and server passes through the RRU (in worker node 1), the backhaul bridge (or TN), and the SPGW (in worker node 3). This scenario can be considered as a Cloud Computing (CC) scenario, because, in reality, worker node 3 may be placed in a data center accessed through an IP network (emulated by the backhaul bridge in this scenario). On the other hand, scenario 2 (Fig.~\ref{fig:demo2p2}) depicts the case where the (NGINX) server and the SPGW are moved to worker node 1 beside the eNB. As a result, this scenario resembles Edge Computing (EC).

Table~\ref{vnft1} shows the performance of the network in terms of the achievable bit rate (using the wget utility in the UE to download an MP4 video from the NGNIX server) for different settings of the bridge. It is observed that the parameters of the transport network highly affect the performance of the system in the CC scenario. For instance, the bit rate is almost halved when the bandwidth is reduced from 10 Mb/s to 5 Mb/s. However, in the EC scenario, the bit rate is independent of variations in the transport network. Furthermore, Table~\ref{vnft2} represents the Round Trip Time (RTT) to an external server (i.e., Google's DNS server at ``8.8.8.8'') for both CC and EC scenarios. In this experiment, the Ethernet switch is connected to the Internet. Therefore, in the EC scenario, ICMP packets to Google's DNS server do not go across the TN bridge, and subsequently, the delay applied to this bridge does not affect the RTT. Although additional delay contributes to the overall RTT in the CC scenario, the relationship between delay and RTT is not flat due to the instability of the external network.

It is worth noting that the LTE bandwidth in our setup is constrained by the USRP bandwidth, which acts as a limiting factor that affects the results in this paper.

\begin{table}[t]
	\begin{center}
		\caption{The effect of TN's bandwidth on the bit rate in different VNF placement scenarios}
		\label{vnft1}
		\begin{tabular}{| c | c | c | c |} 
			\hline
			\multicolumn{2}{| c |}{Transport Network Parameters} & \multicolumn{1}{c}{Scenario 1 (CC)}&\multicolumn{1}{| c |}{Scenario 2 (EC)}\\
			\hline
			Bandwidth (Mb/s) & Delay (ms) & Bit Rate (Mb/s) & Bit Rate (Mb/s) \\
			\hline
			10 & 0 & 1.9 & 1.9 \\
			\hline
			5 & 0 & 0.52 & 1.9 \\
			\hline 
		\end{tabular}
	\end{center}
\end{table}

\begin{table}[t]
	\begin{center}
		\caption{The effect of TN's delay on the RTT in VNF placement scenarios}
		\label{vnft2}
		\begin{tabular}{| c | c | c | c |} 
			\hline
			\multicolumn{2}{| c |}{Transport Network Parameters} & \multicolumn{1}{c}{Scenario 1 (CC)}&\multicolumn{1}{| c |}{Scenario 2 (EC)}\\
			\hline
			Bandwidth (Mb/s) & Delay (ms) & RTT (ms) & RTT (ms) \\
			\hline
			10 & 0 & 120 & 120 \\
			\hline
			10 & 50 & 340 & 120 \\
			\hline 
		\end{tabular}
	\end{center}
\end{table}

\subsection{RAN Slicing}
\begin{figure}[t]
	\centering{\includegraphics[width=0.48\textwidth]{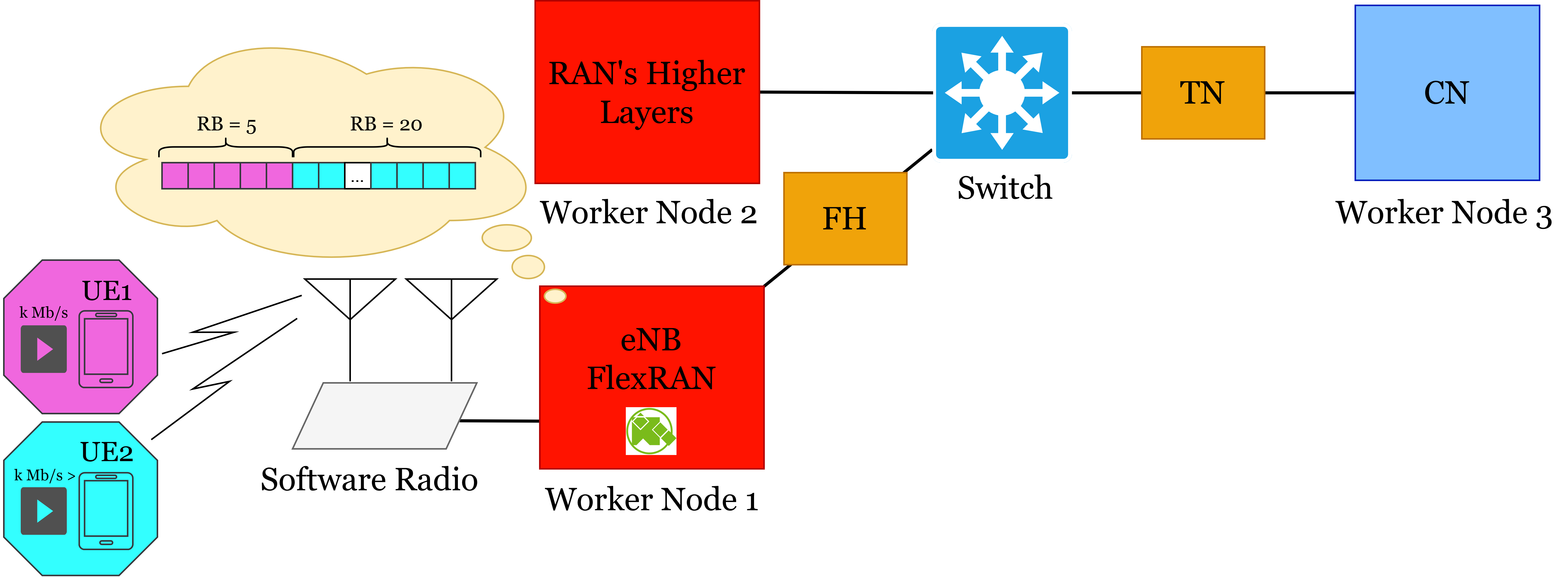}}
	\caption{Test case for the RAN slicing \label{fig:ranslicing}}
 \vspace{-1em}
\end{figure}

The idea of RAN slicing is to assign different numbers of Resource Blocks (RBs) to different network slices in order to comply with their corresponding Service-Level Agreements (SLAs) relevant to the access network. Using FlexRAN along with eNB, we can test the effect of RAN slicing on the performance of the system (Fig.~\ref{fig:ranslicing}). In particular, Table~\ref{rantab} shows the result when we have two slices and one UE in each slice. The total number of RBs is 25 which is divided between slice 1 (UE 1) and slice 2 (UE 2). As expected, by allocating more RBs to each slice, the bit rate of its UE increases. Specifically, the bit rate of UE 1 is 1.05 Mb/s when slice 1 has 5 RBs, and it climbs up to 3 Mb/s as the number of RBs increases to 15.

\section{Related Work}\label{Sec5}
Testbed-based evaluation in wireless networks is a prevalent approach used to assess realistic scenarios \cite{esmaeily2021small}. The importance of testbeds even grows when we come to 5G/B5G mobile networks due to the complexity and different requirements of these networks. This section briefly mentions cellular network testbeds previously developed and their capabilities. 

Rostami et al. \cite{ref10} focused on the orchestration of the TN and RAN by suggesting a hierarchical cross-domain orchestrator that offers network programmability and flexibility. This orchestrator monitors the radio resources at the network access edge level, the transport resources at the access and aggregation levels across multiple domains, and the cloud resources at the network core level to make decisions. The authors demonstrated the advantages and feasibility of their proposed orchestration by implementing two use cases of SDN-based transport and RAN orchestration in a testbed. The first use case presents the sharing of joint RAN-transport resources between two Service Providers (SPs), and the second one demonstrates how an SP can customize its own slice.

Muñoz et al. \cite{ref11} presented the architecture and results of the ADRENALINE as a testbed, which is an SDN/NFV packet/optical transport network and edge/core cloud platform for end-to-end 5G and IoT services, deployed with open-source software and Commercial Off The Shelf (COTS) hardware. Similarly, Fichera et al. \cite{ref12} provided an experimental setup of a convergent 5G service scenario involving IoT, cloud, and edge networks, all featured by SDN capabilities. The implemented testbed also includes an SDN-based orchestrator able to dynamically adapt data delivery paths based on the current load of network switches and links. Another testbed presented by Ramantas et al. \cite{ref13}, which employs COTS components to embody an end-to-end 5G platform based on the C-RAN architecture, with a fully virtualized RAN, an optical/wireless fronthaul, and a cloud-based backend. These approaches do not offer a complete end-to-end network slicing and the source codes needed to deploy the testbeds are not publicly available.

A group of testbeds concentrates on Management and Orchestration (MANO) implementation. For example, BlueArch \cite{ref15} is a 5G testbed providing a hybrid platform for conducting various experiments with different modes of tests, including simulation, emulation, and interaction with the physical network and remote testbed platforms. BlueArch supports ETSI MANO orchestration. The Open MANO and RIFT.io orchestrators are hosted as VMs within a XEN environment. Simula testbed \cite{ref16} also implements a mobile network based on OAI-EPC deployed as a VNF using Open-Source MANO (OSM), which is integrated with C-RAN architecture with functional split capability for BBU processing functions. Likewise, Vittal et al. \cite{ref22} proposed an emulation framework for zero-touch 5G core network slicing management and orchestration that features closed-loop automation. Their framework relies on the use of OSM for NFV MANO functions with NFV orchestration and VNF management functionalities communicating with different Virtual Infrastructure Management (VIM)s.

\begin{table}[t]
	\begin{center}
		\caption{RAN slicing results}
		\label{rantab}
		\begin{tabular}{| c | c | c | c |} 
            \hline
            \multicolumn{3}{| c |}{Scenario} & \multirow{2}{*}{Bit Rate (Mb/s)}\\
			\cline{1-3}
			  No. & Resource Blocks & Device &\\
			\hline
			\rowcolor{yellow} & 5 & UE 1 & 1.05\\
			
			\rowcolor{yellow} \multirow{-2}{*}{1} & 20 & UE 2 & 2.85\\
			\hline
			\rowcolor{pink} & 10 & UE 1 & 1.40\\
			
			\rowcolor{pink} \multirow{-2}{*}{2} & 15 & UE 2 & 1.95\\
			\hline
			\rowcolor{green} & 15 & UE 1 & 3.00\\
			
			\rowcolor{green} \multirow{-2}{*}{3} & 10 & UE 2 & 0.50\\
			\hline
		\end{tabular}
	\end{center}
\end{table}

On the other hand, the testbed proposed by Shorov et al. \cite{ref17} implements end-to-end network slicing. However, it does not offer MANO capability, multi-RATs, and multi-tenancy facilities in the architecture. This testbed utilizes OAI for both RAN and CN domains. There are two CNs that share the radio resources of a single eNB in the RAN. The testbed has been appraised for connection establishment for both normal LTE UE and the one with an implemented Network Slice Selection Assistance Information (NSSAI).

A few works also provide testbed prototypes for research and development activities. ComNetsEmu \cite{10.1109/MCOM.001.2000578} is built as a standalone virtual machine that combines an SDN network emulator (Mininet) with an NFV infrastructure (NFVI) solution (docker) into an integrated framework. However, this is a prototype-only testbed in which deploying any realistic VNFs is left to users.

Some other testbeds are presented by researchers with the objective of AI workload orchestration to facilitate the deployment of AI agents into the testbed. In \cite{ref4}, Connected AI (CAI) is presented as a 5G mobile network testbed with a virtualized and orchestrated structure using containers. CAI focuses on integrating Artificial Intelligence (AI) applications using the Kubeflow tool, albeit it implements partial network slicing and does not incorporate MANO components. It also presents an emulated TN enabling the deployment of any network topology on fronthaul and backhaul, without needing access to actual transport network topologies.

In this paper, we presented \nm{} as an end-to-end, cloud-native cellular network framework, alongside a detailed description of how to build it. In addition, we provided a GitHub repository along with an \textit{Ansible playbook}, which helps researchers and the industrial community to set up \nm{} on their systems to test and evaluate realistic network scenarios. We implemented VNF placement to showcase two different scenarios, EC and CC. RAN slicing has been also implemented as another use case for new-generation cellular networks to demonstrate how efficient allocation of resources can improve the QoS for applications based on their demands. This framework takes advantage of SDN and NFV capabilities, and as a key feature, its main structure is independent of specific VNFs. We provided readers with the required steps for creating a Kubernetes cluster over which the whole cellular network is deployed. Thus, one can create proper definition files for customized VNFs (e.g., 5G SBA core) to deploy our cloud-native testbed for an intended cellular generation. To summarize, Table~\ref{tab3} provides a comparison between cellular network testbeds in the wild. 

\begin{table*}[t]
	\begin{center}
		\caption{Cellular network testbeds and frameworks}
		\label{tab3}
		\begin{tabular}{| c | c | c | c | c | c | c | c | c |} 
			\hline
			Testbed & SDN & NFV & E2E Slicing & MANO & Open-Source & ML-Enabled\\
			\hline
			RAN-transport orchestration testbed \cite{ref10} & \cmark & \cmark & \xmark & \xmark & \xmark & \xmark\\
			\hline
            ADRENALINE testbed \cite{ref11} & \cmark & \cmark & \xmark & \xmark & \xmark & \xmark\\
			\hline
            5G Operating Platform \cite{ref12} & \cmark & \cmark & \xmark & \xmark & \xmark & \xmark\\
			\hline
            C-RAN based 5G platform \cite{ref13} & \cmark & \cmark & \xmark & \xmark & \xmark & \xmark\\
			\hline
			CAI testbed \cite{ref4} & \cmark & \cmark & \xmark & \xmark & \cmark & \cmark\\
			\hline 
			BlueArch \cite{ref15} & \cmark & \cmark & \xmark & \cmark & \xmark & \xmark\\
			\hline 
			SimulaMet OAI EPC testbed \cite{ref16} & \cmark & \cmark & \xmark & \cmark & \cmark & \xmark\\
			\hline 
			The 5G testbed \cite{ref17} & \xmark & \cmark & \cmark & \xmark & \cmark & \xmark\\
			\hline 
			Zero-touch emulation framework \cite{ref22} & \xmark & \cmark & \cmark & \cmark & \xmark & \xmark\\
			\hline
			ComNetsEmu \cite{10.1109/MCOM.001.2000578} & \cmark & \cmark & \xmark & \xmark & \cmark & \xmark\\
			\hline
			Our testbed (\nm) & \cmark & \cmark & \xmark & \xmark & \cmark & \xmark\\
			\hline 
		\end{tabular}
	\end{center}
\end{table*}

\section{Conclusion}\label{Sec6}
Cellular networks have evolved into fully virtualized and programmable networks. As a result, an innovative solution can find its path into operation as easily as some software updates in the network's components. The same virtualization nature of the modern cellular networks also enables building laboratory testbeds for the research and development teams to discover new services/products in isolated, yet close-to-field, environments. In this paper, we shared our findings and best practices in building such a testbed, called \nm{}, for modern cellular networks using state-of-the-art technologies, such as Docker, Kubernetes, ONOS, and Mininet. We especially focused on how to set up a cluster of nodes hosting the cellular network's VNFs and the management entities, and bridges emulating the intermediate IP networks between different parts of a real-world cellular network. Thereby, \nm{} is capable of deploying and testing various scenarios, such as RAN splitting/slicing, edge computing, and VNF placement. Moreover, for a particular open-source project, we walked through the process of installing the required packages, building Docker images and containers, creating pods, setting the configuration files, and deploying the cellular network's core and RAN VNFs on the \nm{}. Finally, the performance of the deployed network was further measured under various test case scenarios to evaluate the benefits of edge computing and RAN slicing.

\noindent
\textbf{Author Contributions.} Conceptualization: Mohammad Reza Heidarpour, Zeinab Zali, Ali Rakhshan, Mahsa Faraji Shoyari; Methodology: Sepehr Ganji, Shirin Behnaminia, Ali Ahangarpour, Erfan Mazaheri, Mohammad Reza Heidarpour, Zeinab Zali; Formal analysis and investigation: Sepehr Ganji, Shirin Behnaminia, Ali Ahangarpour, Erfan Mazaheri, Mohammad Reza Heidarpour, Zeinab Zali; Writing - original draft preparation:  Sara Baradaran, Sepehr Ganji, Shirin Behnaminia, Ali Ahangarpour, Mohammad Reza Heidarpour; Writing - review and editing: Sara Baradaran, Shirin Behnaminia, Sepehr Ganji, Mohammad Reza Heidarpour, Zeinab Zali, Ali Rakhshan, Mahsa Faraji Shoyari; Supervision: Mohammad Reza Heidarpour, Zeinab Zali, Ali Rakhshan, Mahsa Faraji Shoyari.

\noindent
\textbf{Funding.} This study is funded by Mobile Communications Company Research and Development Center, Tehran, Iran.

\noindent
\textbf{Data and Code Availability.} The source codes alongside scripts to set up the framework and execute the use cases are publicly available on our GitHub repository as referenced in the paper.

\section*{Declarations}

\noindent
\textbf{Conflict of Interest.} The authors declare that they have no conflict of interest. 

\noindent
\textbf{Ethical Approval.} Not applicable.

\flushend

\bibliographystyle{IEEEtran} 

\bibliography{References}   

\end{document}